\documentclass[journal=jpclcd,manuscript=letter,layout=onecolumn]{achemso}
\usepackage{graphicx}% Include figure files
\usepackage{dcolumn}% Align table columns on decimal point
\usepackage{bm}% bold math
\usepackage[utf8]{inputenc}
\usepackage[T1]{fontenc}

\usepackage{mathptmx}
\usepackage{siunitx}
\usepackage{graphicx}
\usepackage{dcolumn}
\usepackage{braket}
\usepackage{xfrac}
\usepackage{hyperref}
\usepackage{tikz-cd}
\usepackage{float}
\usepackage{amsmath}
\usepackage[nameinlink,capitalise]{cleveref}
\usepackage{mathtools}

\usepackage{bm}
\DeclareSIUnit\gauss{G}
\DeclareSIUnit{\au}{{a.u.}}
\hypersetup{
colorlinks=true,
linkcolor=blue,
citecolor=blue,
filecolor=green,
urlcolor=blue,
}

\newcommand*{\citen}[1]{%
  \begingroup
    \romannumeral-`\x % remove space at the beginning of \setcitestyle
    \setcitestyle{numbers}%
    \cite{#1}%
  \endgroup   
}

\makeatletter
\renewcommand*{\acs@tocentry@print@aux}{%
  \begingroup
    \let\@startsection\acs@startsection@orig
    \acs@section*{\tocentryname}%
    \tocsize
    \sffamily
    \singlespacing
    \begin{center}
          \begin{minipage}{\acs@tocentry@height}
            \vbox to \acs@tocentry@width{\acs@tocentry@text}%
          \end{minipage}%
    \end{center}%
  \endgroup
}
\makeatother

\title{\Large Signatures of Non-universal Quantum Dynamics of Ultracold Chemical Reactions of Polar Alkali-dimer Molecules with Alkali-metal Atoms:   \\ Li($^2$S) ~+~NaLi($a^3\Sigma^+$) $\to$ Na($^2$S) ~+~Li$_2$({$a^3\Sigma_u^+$})}
\author{Masato Morita$^{1}$, Brian K. Kendrick$^{2}$ {Jacek K{\l}os$^{3,4}$, \\ Svetlana Kotochigova$^{4}$, Paul Brumer$^{1}$$\, ^*$ and Timur V. Tscherbul$^{5}$} }

\affiliation{$^{1}$Chemical Physics Theory Group, Department of Chemistry, and Center for Quantum Information and Quantum Control, University of Toronto, Toronto, Ontario, M5S 3H6, Canada\\
$^{2}$Theoretical Division (T-1, MS B221), Los Alamos National Laboratory, Los Alamos, New Mexico 87545, USA,\\
$^{3}$Joint Quantum Institute, University of Maryland, College Park, Maryland, 20742, USA\\
$^{4}$Department of Physics, Temple University,  Philadelphia, PA 19122, USA\\
$^{5}$Department of Physics, University of Nevada, Reno, NV, 89557, USA}

\date{\today}% It is always \today, today,
\email{paul.brumer@utoronto.ca,ttscherbul@unr.edu}

\begin{document}

\begin{tocentry}
%\vspace{-50pt}
%\includegraphics[scale=0.4, trim = 130 0 0 140 ]{Fig_TOC.pdf}. %temporary
\includegraphics[scale=1.5,  trim = 30 0 0 0]{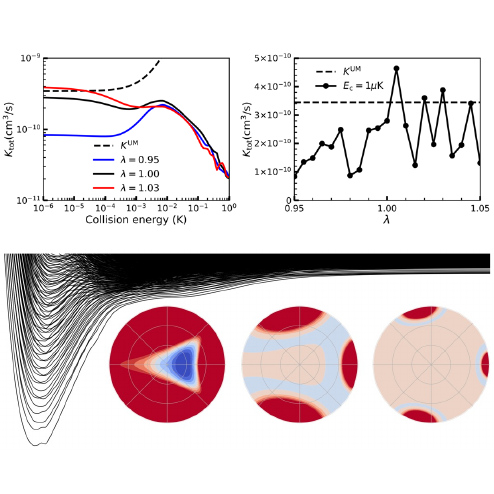}. %temporary
\end{tocentry}

\begin{abstract}

Ultracold chemical reactions of weakly bound triplet-state alkali{-metal} dimer molecules have recently  attracted much experimental interest.
%We use rigorous quantum scattering calculations based on a {newly constructed} {\it ab initio} potential energy surface to explore the chemical  reaction of spin-polarized  NaLi($a^3\Sigma^+$) molecules and Li($^2$S) atoms to form  Li$_2$($a^3\Sigma_u^+$) and Na($^2$S).
We perform rigorous quantum scattering calculations with a new {\it ab initio} potential energy surface to explore the chemical  reaction of spin-polarized NaLi($a^3\Sigma^+$) and Li($^2$S) to form  Li$_2$($a^3\Sigma_u^+$) and Na($^2$S).
The reaction is exothermic, and proceeds readily at ultralow temperatures.
Significantly, we observe strong sensitivity of the total reaction rate to small variations of the three-body part of the Li$_2$Na interaction at short range, which 
we attribute  to a relatively small number of open 
Li$_2$($a^3\Sigma_u^+$) product channels populated in the reaction.
This provides the first signature of highly non-universal dynamics  seen in rigorous quantum reactive scattering calculations of an ultracold exothermic insertion reaction 
involving a polar alkali-dimer molecule, opening up the possibility of  probing microscopic interactions in atom+molecule collision complexes via ultracold reactive scattering experiments.

\end{abstract}

\maketitle

%{\it Introduction.} {$[ \leftarrow$ to be deleted before JPCL submission. ]}\\
The study of chemical reactions at low and ultralow temperatures is projected to create major advances in our understanding of chemical reactivity at the most fundamental level \cite{Carr:09,Herschbach:09,Balakrishnan:16, Bohn:17,Liu:22}.  
The large deBroglie wavelengh of ultracold molecular reactants \cite{Herschbach:09} combined with the experimenter's ability to prepare them in single well-defined quantum states give rise to pronounced  quantum effects, making ultracold chemical reactions an ideal platform for exploring the impact of these fascinating effects on chemical reactivity.
Examples include threshold and resonance scattering \cite{Balakrishnan:16,Tscherbul:15,Hermsmeier:21,Son:22}, quantum reflection \cite{Orzel:99,Mody:01,Bai:19},  quantum statistics \cite{Ni:10,Quemener:12}, and quantum coherent control \cite{Devolder:20,Devolder:21}.
 In addition,  the detection  of reaction products and intermediate collision complexes has been experimentally realized \cite{Hu:19,Liu:20}, enabling precision tests of  long-held statistical theories of chemical  dynamics \cite{Liu:21}.

%Most of the previous experimental and theoretical studies  have focused on ultracold reactions of either covalently bound molecules (such as ground-state alkali-{metal} dimers  KRb and NaK) or extremely weakly bound Feshbach molecules   \cite{Kohler:06,Rui:17}.
  
All chemical reactions studied thus far at low and ultralow temperatures
can be classified into abstraction and insertion types \cite{Weck:06,Krems:19}. Abstraction reactions, such as F~+~H$_2$~$\to$ HF~+~H \cite{Balakrishnan:01,Tizniti:14,DeFazio:19} and H(D)~+~H$_2$ $\to$ H$_2$(HD)~+~D \cite{Simbotin:11,Simbotin:15,Kendrick:15,Kendrick:18,Kendrick:19} possess an activation barrier, whereas insertion reactions, such as K~+~KRb $\to  $ K$_2$~+~Rb \cite{Croft:17,Croft:17b} and KRb~+~KRb $\to$ K$_2$~+~Rb$_2$  \cite{Ospelkaus:10,Liu:21}  are barrierless and characterized by a deep potential well  \bibnote{A notable exception is vibrational excitation of the reactant molecule, for which abstraction reactions can also become barrierless and exhibit a potential well (e.g., H(D) + H$_2$($v\geq5$)) (See Refs.~\citen{Simbotin:15,Kendrick:18,Kendrick:19}).}. 
%\cite{comment1}.

%A notable exception is vibrational excitation of the reactant molecule for which abstract reactions
%can also become barrierless and exhibit a potential well (e.g., D + H$_2$($v\geq5$)) \cite{Simbotin:11,Simbotin:15}.
 Ultracold insertion reactions are generally well described by universal models (UMs)  \cite{Quemener:12,Quemener:10,Kotochigova:10,Idziaszek:10, Croft:17,Tscherbul:20,Liu:21},  
 which assume that once  the reactants approach each other at close range, they react with unit probability.  
 Exact quantum dynamics calculations showed that the rate of the 
{K~+~KRb($X^1\Sigma^+$)} 
chemical reaction is in excellent agreement with the UM prediction \cite{Croft:17} and with  experiment \cite{Ospelkaus:10}. 
For other insertion chemical reactions, such as 
{ Li~+~YbLi($X^2\Sigma^+$) $\to$ Li$_2(X^1\Sigma^+_g)$~+~Yb \cite{Makrides:15,Li:19b} and Li~+~NaLi  ($X\,^1\Sigma^+$)  $\to$ Li$_2(X\,^1\Sigma^+_g)$~+~Na  \cite{Kendrick:21,Kendrick:21b},} 
the calculated deviations from the UM do not exceed 30\%.
 
 %Exact quantum dynamics calculations showed that the rate of the K~+~KRb chemical reaction is in perfect agreement with the UM prediction \cite{Croft:17} and with  experiment \cite{Ospelkaus:10}. For other insertion chemical reactions, such as Li~+~YbLi $\to$ Li$_2$~+~Yb and Li~+~NaLi($X$) $\to$ Li$_2(X^1\Sigma^+_g)$~+~Na, the calculated deviations from the UM do not exceed 30\%.

While UMs are an important tool
% which allows one to estimate the rates of ultracold chemical reactions 
\cite{Kotochigova:10,Julienne:11,Idziaszek:10}, they do not provide insight into microscopic interactions within the reaction complex \cite{Tscherbul:15}, 
%nor do they allow control of reaction dynamics via scattering resonances \cite{Tscherbul:15,Frye:15}. 
which is crucial for designing mechanisms to control the reaction dynamics via, e.g., scattering resonances \cite{Tscherbul:15,Frye:15}. Similarly, they provide no information about conditions for the validity of UMs, or the parameter dependence of the dynamics.
Hence, identifying systems showing significant deviations from universal behavior  ({\it non-universal effects}) is an important goal, which can yield insights into short-range interactions in the reaction complex, and enable one to tune ultracold chemical reactivity with scattering resonances, 
a long sought-after goal of ultracold chemistry \cite{Krems:08,Balakrishnan:16}.
Signatures of non-universal  effects were observed experimentally in ultracold collisions of Li$_2$ Feshbach molecules with Li atoms \cite{Wang:13}, and, very recently, in ground rovibrational state molecule-atom and molecule-molecule collisions, such as K~+~NaK($X^1\Sigma^+$) \cite{Yang:19},  Na~+~NaLi($a^3\Sigma^+$)   \cite{Son:22} and  NaLi~+~NaLi  \cite{Park:23}, where the observed reaction rate coefficients deviate from UM predictions by several orders of magnitude. 
The ultracold chemical reaction of polar NaLi($a^3\Sigma^+$) molecules with Li atoms to form the Li$_2$($a^3\Sigma_u^+$) + Na products is a natural candidate to search for non-universal effects, as it is energetically allowed [unlike the Na~+~NaLi($a^3\Sigma^+$) $\to$ Na$_2$($a^3\Sigma_u^+$)~+~Li reaction \cite{Hermsmeier:21}], experimentally feasible \cite{Rvachov:17,Son:20,Son:22}, and  amenable to rigorous theoretical studies due to light reactants. %reaction However, these effects are yet to be observed in rigorous quantum dynamics calculations on ultracold chemical reactions involving a polar alkali-dimer molecule.
%While non-universal effects can be described phenomenologically by introducing a short-range loss parameter $y$ in the UM in the framework of multichannel quantum defect theory \cite{Idziaszek:10}, marked 

Non-universal behavior has not yet been seen in rigorous quantum  scattering calculations on ultracold chemical reactions of polar alkali dimers.
%Ultracold weakly bound alkali dimers NaLi($a^3\Sigma^+$) and Li$_2$($a^3\Sigma^+_u$)    have been produced in their rovibrational ground states  via photoassociation \cite{Rvachov:17,Polovy:20}, and non-universal collisions have been observed in ultracold atom-molecule mixtures \cite{Yang:19,Son:22} as noted above.
 %Sympathetic cooling of trapped  NaLi($a^3\Sigma^+$) molecules with ultracold Na atoms has been demonstrated \cite{Son:20}. 
 %Despite the profound significance of these observations,  
 Such behavior has only been predicted for ultracold reactions of {\it nonpolar} molecules,
 such as Li~+~Li$_2$($a^3\Sigma_u^+)$ \cite{Cvitas:05,Cvitas:07},  Na~+~Na$_2$($a^3\Sigma_u^+)$ \cite{Soldan:02},  K~+~K$_2$($a^3\Sigma_u^+)$ \cite{Quemener:05}, and $^7$Li~+~${^6}\text{Li}^{7}\text{Li}$ $\to$ $^7$Li$_2$~+~${^6}\text{Li}$ \cite{Cvitas:05b}. These reactions are unique in that (i) they have a vanishingly small exothermicity, so only a few product states are  populated at ultralow temperatures \cite{Hutson:07} and (ii) {their reactants and products are identical (apart from isotopic substitution).} As a result, their zero-temperature  rates tend to be below the universal values, on the order of $4.7\times 10^{-12}$ cm$^3$/s for the $^7$Li~+~${^6}\text{Li}^{7}\text{Li}$ reaction \cite{Hutson:07}, and may not be detectable experimentally. 
In the most thoroughly studied cases where the  reactants and products are identical, it is not even possible to disentangle chemical reactivity from inelastic processes \cite{Hutson:07}.
By contrast, ultracold chemical reactions probed in recent experiments  \cite{Son:20,Son:22} involve  {\it polar} triplet-state alkali dimers (such as NaLi), and can be exothermic by several hundreds of Kelvin, potentially populating %tens to hundreds of 
dozens of  product rovibrational states.

Here, we show that pronounced non-universal effects can occur in the ultracold insertion reaction of spin-polarized triplet NaLi molecules and Li atoms, Li($^2$S) + NaLi($a^3\Sigma^+,v=0,\,j=0$) $\to$ Li$_2$($a^3\Sigma_u^+,v',j'$)~+~Na($^2$S), significantly extending the range of systems displaying non-universal behavior.
% which, by virtue of the small binding energy of NaLi($a^3\Sigma^+$), the distinguishability of the reactants and products, and its substantial exothermicity,  serves as a clear-cut  example of an ultracold vdW reaction that could be probed in near-future experiments. 
We use rigorous quantum scattering calculations based on {our}  {\it ab initio} interaction potentials to map out the dependence of the reaction rates on the incident collision energy. 
Our calculations show that the reaction  occurs at a high rate, and hence it can be detected experimentally in an ultracold spin-polarized Li~+~NaLi mixture. 
Further, {\it the total reaction rate 
is sensitive to small variations of the short-range part of the interaction potential even after summing over all state-to-state  reaction rates, and can deviate by a factor of 
$\simeq4$ from the UM prediction}, representing the first rigorous theoretical demonstration of non-universal behavior in an ultracold  insertion %chemical 
reaction involving a polar molecule.  (Note that the non-universal behavior recently observed  in ultracold K~+~NaK, Na~+~NaLi, and NaLi~+~NaLi collisions  \cite{Yang:19,Son:22,Park:23} is not yet amenable to rigorous quantum scattering calculations due to the large density of rovibrational states of the reactants, non-adiabatic effects, and external fields present in these experiments). 
We attribute this highly non-universal behavior to a relatively low density of resonance states in this reaction and  a limited number of open product channels  compared to the chemical reactions explored previously \cite{Croft:17,Makrides:15,Li:19b,Kendrick:21,Kendrick:21b}, which displayed close to universal behavior.
Our results suggest that non-universal effects can be experimentally observed in ultracold chemical reactions of spin-polarized triplet-state alkali-metal dimers with alkali-metal atoms. 
This opens up the prospects of mapping atom-molecule  interactions within the reaction complex, and of controlling the reaction dynamics with external fields, motivating further experimental and theoretical research into {non-universal ultracold chemistry.}

\begin{figure}[t]
\begin{center}
\includegraphics[height=0.34\textheight, keepaspectratio,trim = 0 50 100 100]{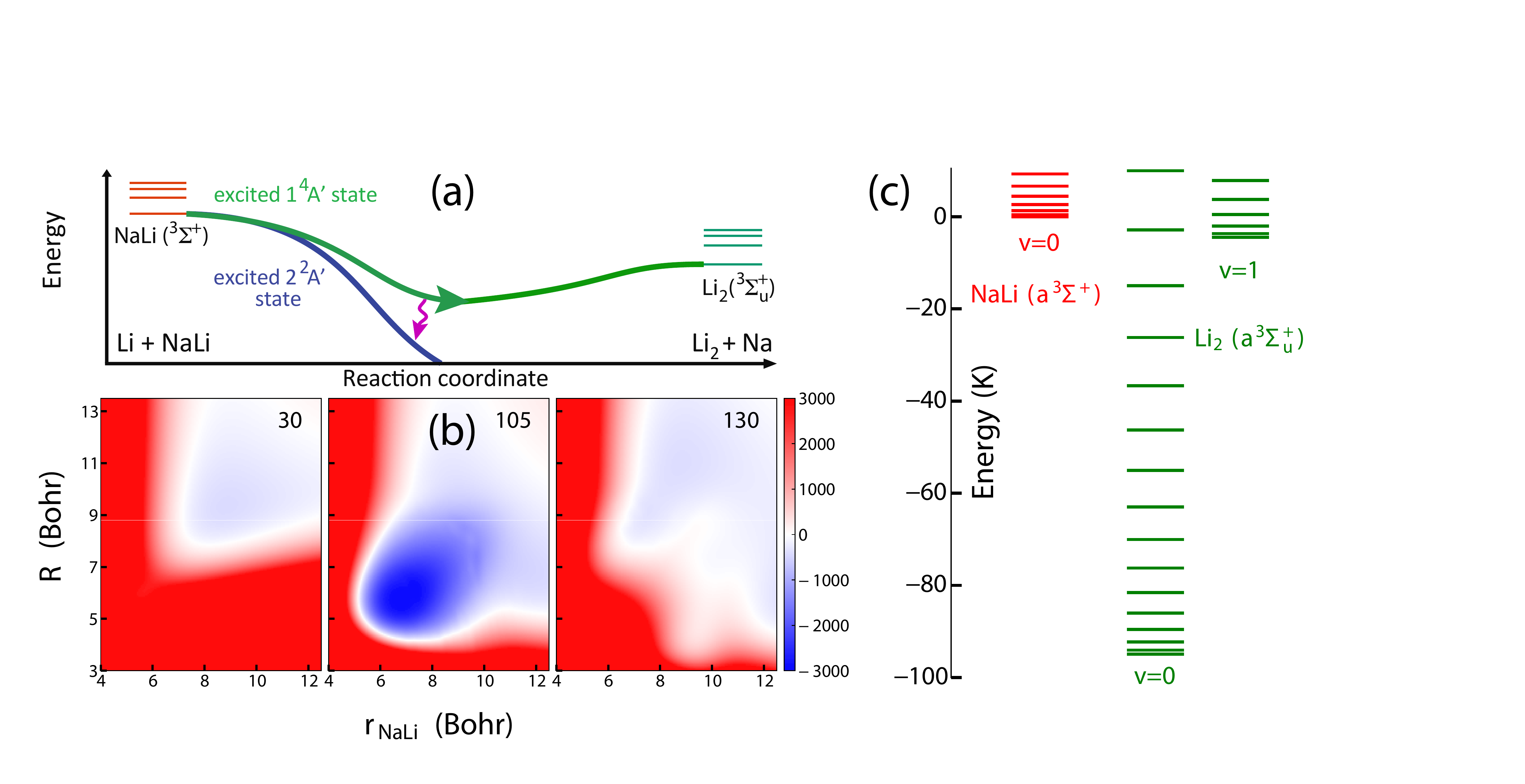}
\end{center}
\caption{(a) Schematic diagram of the chemical reaction Li~+~NaLi$(a^3\Sigma^+)$ $\to$ Li$_2(a^3\Sigma_u^+)$~+~Na. The reaction of spin-polarized reactants proceeds  via a spin-conserving pathway on the high-spin $1^4A'$ PES (green line) as predicted by the Wigner spin rule. The spin-nonconserving reaction pathway  forbidden by the Wigner spin rule (not considered in this work)  requires a non-adiabatic transition to the excited $2^2A'$ PES (wavy  line).    
{(b) {\it Ab initio} PES for the  $1 ^4A^\prime$ electronic state of the Li$_2$Na trimer computed in this work plotted as a function of the Jacobi coordinates ($R$, $r_\mathrm{NaLi}$) at the Jacobi angles $\theta=30^\circ$ (left), $\theta=105^\circ$ (middle), and $\theta=130^\circ$ (right) in the Li~+~NaLi arrangement. 
 (c) Energy diagram for the rovibrational states of the LiNa ($a^3\Sigma^+$) reactant and of the Li$_2$ ($a^3\Sigma^+_u$) product. 
 The zero of energy  corresponds to the energy of the rovibrational ground state of LiNa ($a^3\Sigma^+,v=0,j=0$). }}
\label{fig:diagramPES}
\end{figure}

%\section{Potential energy surface and quantum dynamics of the ultracold Li~+~NaLi $\to$ Li$_2$~+~Na chemical reaction} 
%{\it Quantum dynamics of the ultracold  chemical reaction $\textrm{Li}$~+~$\textrm{NaLi}$ $\to$ $\textrm{Li}_2$~+~$\textrm{Na}$.}  {$[ \leftarrow$ to be deleted before JPCL submission. ]}\\\
We consider the chemical reaction of $^{23}$Na$^6$Li 
molecules in their {metastable triplet electronic states} $a^3\Sigma^+$, created in recent experiments \cite{Rvachov:17}.  
%Unlike ground-state NaLi dimers, triplet NaLi($a^3\Sigma^+$) molecules have unpaired electron spins, and possess XXX of internal (electron) energy, measured relative to the ground-state energy.
\Cref{fig:diagramPES} (a) is a schematic of the reaction path of the Li($^2$S)~+~NaLi($a^3\Sigma^+$) $\to$ Li$_2$($a^3\Sigma_u^+$)~+~Na($^2$S) reaction.
%The reaction path of energetics anof  the Li~+~NaLi chemical reaction is shown schematically in 
The three-fold spin degeneracy of NaLi($a^3\Sigma^+$) and the two-fold spin degeneracy of Li($^2$S) give rise to two adiabatic potential energy surfaces (PESs) of  doublet ($2 ^2A^\prime$, $S=1/2$) and quartet  ($1 ^4A^\prime$, $S=3/2$) symmetries, where   $S$ is the total spin of the reaction complex. The PESs are split at short range by a strong spin-exchange interaction.
While the high-spin $1 ^4A^\prime$ PES is the lowest state in the quartet symmetry,  the doublet $2 ^2A^\prime$ is the {\it first excited state} in the doublet spin symmetry. The ground-state $1 ^2A^\prime$ PES [not shown in \cref{fig:diagramPES} (a)] correlates with the Li($^2$S)~+~NaLi($X^1\Sigma^+$) reactants in their ground electronic states \cite{Kendrick:21}, which lie 9629~K below the Li($^2$S)~+~NaLi($a^3\Sigma^+$) asymptote of interest here.
The ground and the first excited doublet PESs exhibit a conical intersection (CI), which has a pronounced effect on the quantum dynamics of the ultracold chemical reaction of {\it ground-state} reactants Li($^2$S)~+~NaLi($X^1\Sigma^+$) $\to$ Li$_2$($X^1\Sigma_g^+$)~+~Na($^2$S) \cite{Kendrick:21}.

We assume that both NaLi($a^3\Sigma^+$) molecules and Li atoms are prepared in their fully polarized electron spin states prior to the reaction, which is readily achievable experimentally  \cite{Bohn:17,Rvachov:17,Son:22}.
%given the exquisite level of control  over the internal states of ultracold molecules  attained 
Thereby, the reaction complex is initialized in a state with  the total electron spin projection $M_S=\pm 3/2$, which corresponds to $S=3/2$. Following previous theoretical work \cite{Cvitas:05,Cvitas:05b,Hutson:07}, we assume the validity of the Wigner spin rule \cite{Moore:73,Tscherbul:06,Krems:08,Hermsmeier:21,Haze:22}, which states that $S$, the total electron spin of the reaction complex, is conserved. This allows us to restrict attention to the quantum dynamics on a single adiabatic $1^4A^\prime$ PES and neglect the intersystem crossing transitions to the low-spin $1 ^2A^\prime$  and $2 ^2A^\prime$  PESs, which exhibit a CI \cite{Kendrick:21}. The CI could, in principle, affect the dynamics via intersystem crossing transitions mediated by weak $S$-nonconserving couplings between the  $1^4A^\prime$  PESs and the low-lying PESs due to  e.g., the intramolecular spin-spin interaction in NaLi \cite{Abrahamsson:07,Hermsmeier:21}. 
%PESs exhibit a CI \cite{Kendrick:21}, which could affect the dynamics 
Model calculations show that such $S$-nonconserving pathways are typically suppressed by the factor of $>$10 compared to their $S$-conserving counterparts in chemical reactions of light atoms and molecules \cite{Tscherbul:20,Hermsmeier:21}, although infrequent exceptions can occur at certain collision energies and magnetic fields.
%Indeed, model calculations have shown these spin-nonconserving transitions to be several orders of magnitude slower than  
To rigorously quantify the effect of the doublet PESs and their CI, it is necessary to carry out quantum reactive scattering calculations, which  explicitly account for the interactions between the high and low-spin PESs ($1^4A^\prime$,  $1 ^2A^\prime$,  and $2 ^2A^\prime$). At present, this is computationally unfeasible, but progress towards such calculations is currently underway in our laboratories.
%Progress towards such calculations is currently underway in our groups.

We have used the Multi-Reference Configuration Interaction (MRCI) method \cite{knowles:88,werner:88} to calculate the $1^4A^{\prime}$ PES of the NaLi$_2$ trimer based on
reference wave functions for the two lowest spin-polarized states, $1 ^4A^\prime$  and $2 ^4A^\prime$. These reference functions have been obtained from  state-averaged Multi-Reference Self-Consistent-Field (MCSCF) calculations~\cite{knowles:85,werner:85}. The three valence electrons that describe the NaLi$_2$ trimer are correlated using an active space composed of 12 orbitals, where nine and three are of $A^\prime$ and $A^{\prime\prime}$ symmetry, respectively. {The basis sets and effective core and polarization potentials used for Na and Li atoms are described} in detail in our recent work \cite{Kendrick:21}, which was focused on the  energetically lowest spin doublet states $1 ^2A^\prime$  and $2 ^2A^\prime$.  We performed  the  electronic structure calculations using the MOLPRO package~\cite{MOLPRO}, with further details given in the Supporting Information (SI) \cite{SM}.
The PES for the $1^4A^{\prime}$ electronic state of the Li$_2$Na trimer used in our reactive scattering calculations has the form
\begin{equation}\label{pes}
V_{1^4A^{\prime}}(r_a,r_b,\alpha) = V^\text{pairwise}_{1^4A^{\prime}}(r_a,r_b,\alpha) + \lambda V^\text{3-body}_{1^4A^{\prime}}(r_a,r_b,\alpha)
\end{equation}
where $r_a$ and $r_b$ denote the two NaLi bond lengths, $\alpha$ is the bond angle for Li(a)-Na-Li(b), $V^\text{pairwise}_{1^4A^{\prime}}(r_a,r_b,\alpha)$ is the three atom pairwise (two-body) potential obtained by adding together the spectroscopically accurate NaLi($a^3\Sigma^+$) and Li$_2$($a^3\Sigma_u^+$) dimer potentials from Refs.~\cite{Lesiuk:20,Hermsmeier:21} and $V^\text{3-body}_{1^4A^{\prime}}(r_a,r_b,\alpha)$ is the non-additive three-body contribution\cite{SM}. The scaling parameter $\lambda$ is used below to explore the sensitivity of our results to small variations in the non-additive three-body part of the PES.
Unless stated otherwise, we assume $\lambda=1$.

%\Cref{fig:diagram} (a) shows a schematic diagram of the potential energy surfaces (PESs) for the triatomic Li$_2$Na complex associated with the NaLi($a^3\Sigma^+$)+Li($^2$S) $\to$ Na($^2$S)+Li$_2$($a^3\Sigma^+_u$) reactions. The doublet and triplet collision partners give rise to the quartet and doublet PESs. 

%Our concern in this paper is the reaction of the fully spin-polarized NaLi($a^3\Sigma^+$) and Li($^2$S), which proceeds via a spin-conserving pathway on the $1 ^4A^\prime$ PES because the spin-nonconserving reaction pathway via the $2 ^2A^\prime$ state  is forbidden by the Wigner spin rule \cite{Moore:73}.  In fact, a recent study with model statistical calculations demonstrated that the spin-nonconserving pathways is suppressed by a factor of 10-100 for a reaction with a similar system of Na+NaLi \cite{Hermsmeier:21}.

\Cref{fig:diagramPES}(b) shows contour plots of our {\it ab initio} Li$_2$Na PES as a function of the  Jacobi coordinates ($R$, $r_\mathrm{NaLi}$ ) for several value of Jacobi angles in the Li~+~NaLi arrangement. The PES is barrierless and the global minimum occurs at $\theta=105^\circ$ in the Li~+~NaLi arrangement ($R=5.75$~$a_0$ and $r=6.85$ $a_0$) as shown in the middle panel of \cref{fig:diagramPES}(b).
The PES depth counted from the minimum of the Li$_2$ ($a^3\Sigma^+_u$) product's potential energy curve is 3012~K.  
The  exothermicity of the reaction Li~+~NaLi($a^3\Sigma^+,v=0,j=0$) $\to$ Li$_2$($a^3\Sigma_u^+,v’=0,j’=0$)~+~Na is $\sim$94.9~K ($\sim$112.0~K) including (excluding) the zero-point energies of the reactants and products. This makes it possible for the reaction to occur  at  ultralow collision energies. \Cref{fig:diagramPES}(c) shows the internal rovibrational energy levels  of the reactants and  products calculated from the accurate {\it ab initio} potential energy curves of NaLi($a^3\Sigma^+$) and Li$_2$($a^3\Sigma_u^+$) \cite{Lesiuk:20,Hermsmeier:21}. In calculating the energy levels, we neglected the small splittings due to the fine and hyperfine structure of the reactants and products, as done in prior theoretical work on nonpolar alkali dimers \cite{Cvitas:05,Cvitas:05b,Hutson:07}. This is expected to be a reasonable approximation for fully spin-polarized reactants \cite{Hutson:07}. In the limit of zero collision energy, a total of 18 rovibrational states of Li$_2$ are energetically accessible, including 15 states in the $v'=0$ manifold  ($j'=0-14$) and 3 states in the  $v'=1$ manifold ($j'=0-2$).

%NaLi ($a^3\Sigma^+$) and product Li$_2$ ($a^3\Sigma^+_u$) based

%{\Cref{fig:diagram} (a) shows a schematic diagram for the potential energy surfaces of the triatomic Li$_2$Na complex associated with the NaLi($a^3\Sigma^+$)+Li($^2$S) $\to$ Na($^2$S)+Li$_2$($a^3\Sigma^+_u$) reactions. 
% of the quarter and doublet  PESs  doublet and triplet collision partners give rise to the quartet and doublet potential energy surfaces for the collision complex. 

%{The behaviour of the  $1 ^4A^\prime$ along with the collinear geometry is shown in \Cref{fig:diagram} (b). The arrows in the figure denotes the entrance and exit channel of the reaction. There is a relatively shallow well in the PES and no reaction barrier. The reaction is exothermic (exothermicity $\sim$ 90 Kelvin)
% {The energy diagram for the internal rovibrational states of the  LiNa ($a^3\Sigma^+$) and Li$_2$ ($a^3\Sigma^+_u$) is shown in the panel (c). The internal energy of the reactant, Li+LiNa($v=0,j=0$), is set to the energy of zero. At the zero collision energy limit, 18 rovibrational states for Li$_2$ , $j'=0-14$ and $j'=0-2$ for the $v'=0$ and $v'=1$ manifolds, respectively, are energetically accessible.}

\begin{figure}[t]
\begin{center}
\includegraphics[height=0.44\textheight,keepaspectratio]{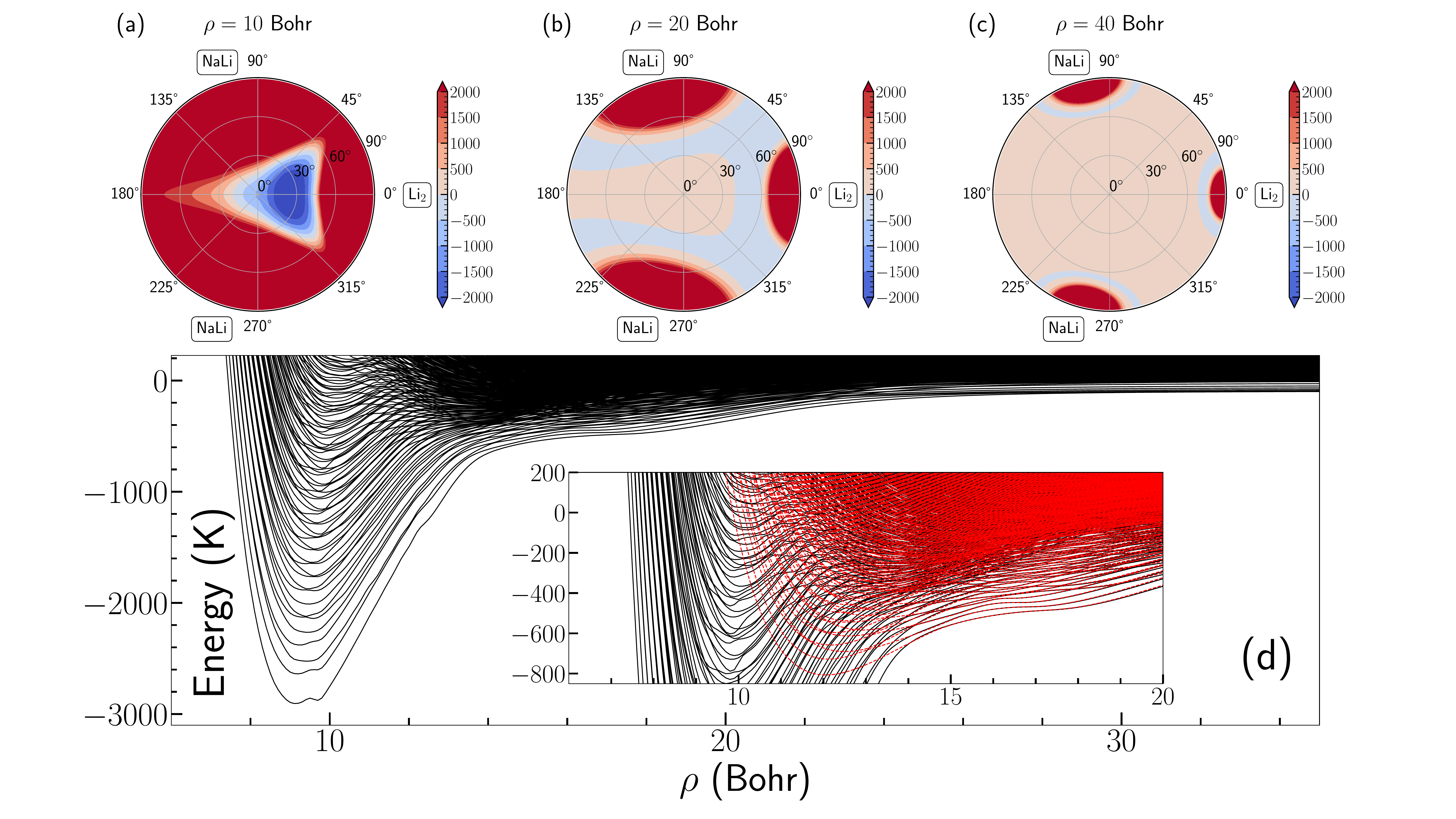}
\end{center}
\caption{
(a)-(c) Stereographic projections of the {\it ab initio} $1 ^4A'$  PES of the NaLi$_2$ trimer for several values of hyperradius (in $a_0$): 
{$\rho=10$ } 
(a), $\rho=20$ (b), and $\rho=40$ (c).  The north pole of the hypersphere is centered at the origin and the zero of energy  is chosen at the rovibrational ground state of the NaLi reactant. The region of configuration space accessible  at ultralow energies is indicated with blue color. (d) The adiabatic potential curves as a function of $\rho$ for $J=0^+$ and even exchange symmetry of Li nuclei.  The insert shows the adiabatic curves in the vicinity of the reactant threshold. The adiabats  calculated without the three-body interaction are shown by the red (light grey) dashed lines.
}
\label{fig:adiabats}
\end{figure}

To study the quantum dynamics of the ultracold Li~+~NaLi $\to$ Li$_2$~+~Na reaction, we use a rigorous quantum dynamics approach based on the adiabatically adjusting principal-axis frame hyperspherical (APH) coordinates  as described in SI \cite{SM}. 
\Cref{fig:adiabats}(d) shows the adiabatic eigenvalues $\epsilon^{Jpq}_{n}(\rho_{\xi})$  of  the Li$_2$Na reaction complex
calculated  using the accurate {\it ab initio} PES (see SI \cite{SM} for technical details).  The corresponding fixed-$\rho$ contour plots of the $1 ^4A'$ PES as a function of the polar and azimuthal hyperangles are displayed in panels (a)-(c). The entire configuration space can be divided up into three distinct regions. In the short-range region ($\rho=8-15$ $a_0$) a strongly bound NaLi$_2$ reaction complex forms, and the different reaction arrangements are strongly coupled. The adiabatic energies range from -3,000 K (near the PES minimum) up to the three-body breakup threshold and above.

In the intermediate region ($\rho=15-40$ $a_0$) the reactant and product arrangements, while still strongly coupled, begin to separate from one another. As shown in \cref{fig:adiabats} (c), at $\rho=20\,a_0$ there still exists a large region of configuration space between the reactant  and product reaction arrangements (shown in light blue color) where  the PES is large and negative.
% Specifically, narrow reaction paths existing in the collinear configuration of the entrance channel are energetically accessible to the  product arrangement through paths near the polar angle $\theta=90 ^\circ$.
Thus,  the reactive scattering wavefunction at $\rho=20\,a_0$ is still substantially delocalized between the reactant and product arrangements.
 Thus is because both the reactants and products of the Li~+~NaLi($a^3\Sigma^+$) $\to$ Li$_2$($a^3\Sigma^+_u$)~+~Na  chemical reaction are weakly bound (van der Waals, vdW) molecules held together by long-range dispersion forces, with binding energies  not exceeding 300~K \cite{Parker:02,Gronowski:20}. 
 {\it This makes the vdW chemical reaction studied here significantly different from those of covalently bound molecules} such as Li~+~NaLi($X^1\Sigma^+$)$\to$ Li$_2$($X^1\Sigma_g^+$)~+~Na \cite{Kendrick:21,Kendrick:21b}, where  the separation of the reactant and product  arrangements is essentially complete at much smaller $\rho\simeq 10-15\,a_0$.  This persistence of reactive (inter-arrangement) couplings up to a large $\rho$ is due to the proximity to the three-body breakup threshold, and may be considered as a distinctive feature of quantum dynamics of vdW chemical reactions. We note that in three-body recombination reactions, which start above the three-body threshold, such inter-arrangement couplings occur even asymptotically at large $\rho$ \cite{DIncao:18}. Ultracold vdW reactions thus occupy an intermediate position between conventional chemical reactions, which take place at very close range ($\rho \leq15\, a_0$), and three-body recombination transitions (or chemical reactions of Feshbach molecules), which occur  at very long range ($\rho \geq 100\, a_0$).

The value of $\rho=40$ $a_0$ marks the beginning of the long-range region,   where the entrance and exit reaction arrangements  are completely separated on the surface of a hypersphere by a barrier with a height of $>200$~K (note that this barrier does not correspond to an actual barrier in the reaction path). In the limit $\rho\to\infty$, different reaction arrangements become localized at specific hyperangles $\theta_{\alpha}$, which correspond to the minima of  the diatomic molecule's potentials in each arrangement.
The adiabatic curves computed without the three-body terms ($\lambda=0$) are nearly identical to their three-body counterparts already at  $\rho\ge 14\, a_0$ as shown in the inset of \Cref{fig:adiabats}(d). The three-body terms dominate at short range  due to  a  conical intersection between the ground and the first excited quartet PESs \cite{Colavecchia:03,Hutson:07,SM}.

\begin{figure}[t]
\begin{center}
\includegraphics[height=0.45\textheight,keepaspectratio,trim = 0 40 0 0]{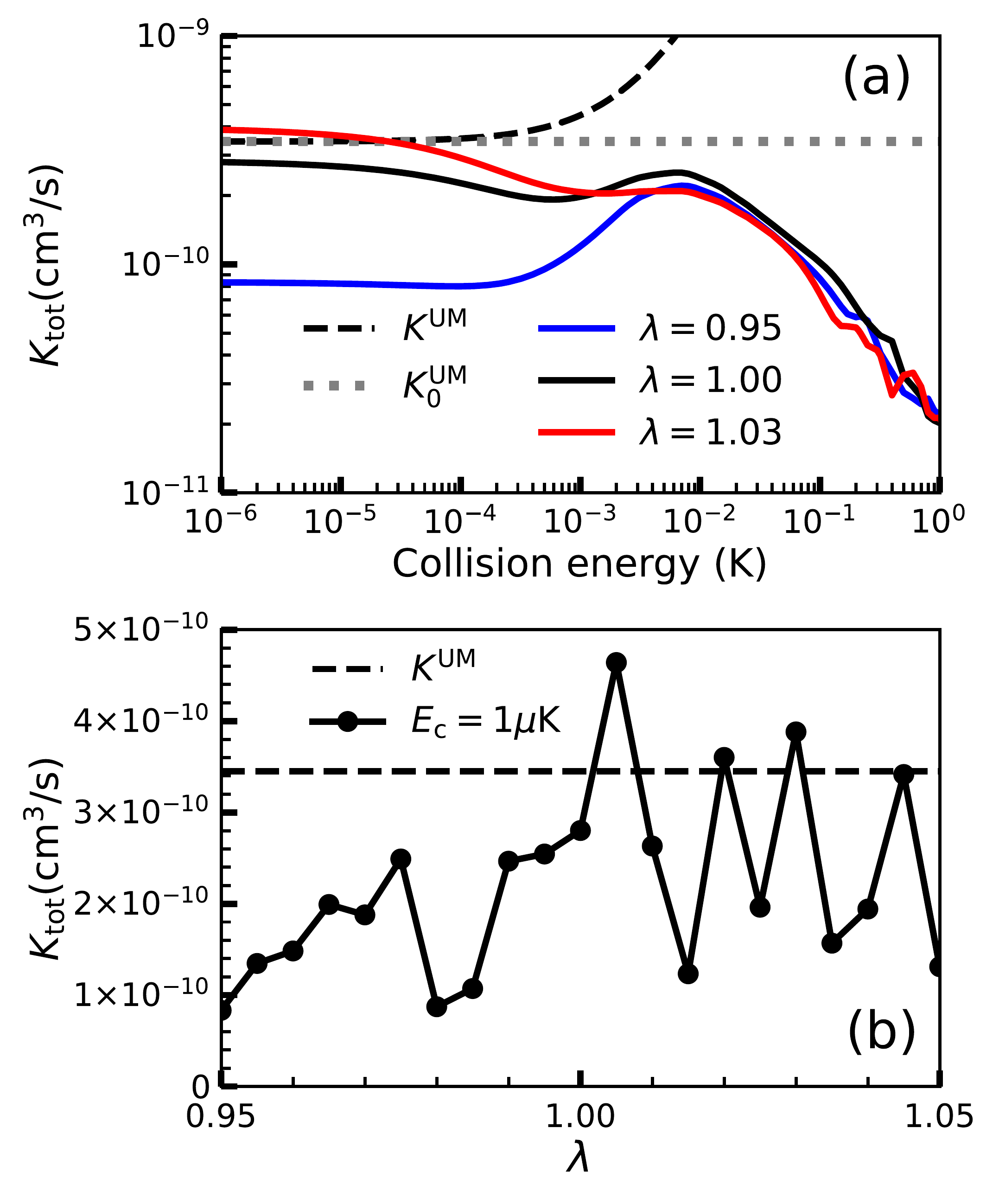}
\end{center}
\caption{(a) Total rate coefficients for the Li~+~NaLi($a^3\Sigma^+,v=0,j=0$) $\to$ Li$_2$($a^3\Sigma^+_u$)~+~Na chemical reaction plotted as functions of collision energy for different values of the PES scaling parameter $\lambda$. The universal rate $K^\text{UM}$ (long dashed line) and its s-wave component $K_0^\text{UM}$ (short dashed line) are shown.  (b) Same as panel (a) at $E=1\, \mu$K  plotted as a function of $\lambda$.  }
\label{fig:rates}
\end{figure}

\Cref{fig:rates}(a) shows the total rate of the  Li~+NaLi($a^3\Sigma^+$) chemical reaction, $K_\text{tot}=\sigma_r v$, where 
$\sigma_r$ is the total reaction cross section, and
$v=\hbar k/\mu$ is the collision velocity, calculated for several values of the scaling parameter $\lambda$. The rates approaches a constant value in the $s$-wave limit of zero collision energy in keeping with the Wigner threshold law ($K_\text{tot}\simeq k^{2l}$). 
%of zero collision energy, whereas the $p$-wave rate (or the $J=1$ rate for the $j=0$ incident channel) scales as $\simeq k^{2}$
 Also plotted in \Cref{fig:rates}(a) is the total universal reaction rate calculated as $K^\text{UM} = K_0^\text{UM} + K_1^\text{UM}(E)$, where $K_0^\text{UM}$ and $ K_1^\text{UM}(E)$ are the $s$-wave and $p$-wave  universal rates calculated as described in Refs.~\citen{Idziaszek:10,Julienne:11}. These are evaluated under the assumption that all of the incoming flux is fully absorbed at short range (loss parameter $y=1$), which leads to the same universal rate expression,
parametrized by the long-range dispersion coefficient $C_6$ and the reduced mass for the collision,
regardless of the details of  short-range interactions \cite{Idziaszek:10,Julienne:11} (see the SI \cite{SM} for more details).

To obtain the universal rates, we calculated the value of $C_6=2891$ a.u. for the Li-NaLi($a^3\Sigma^+$) trimer using a high-level spin-restricted coupled-cluster method with single, double, and perturbative triple excitations RCCSD(T) and an aug-cc-pwCVQZ basis set with a bond length of NaLi at 9.0906~$a_0$ corresponding to the average internuclear  distance of NaLi($a^3\Sigma^+,v=0,j=0$) \cite{SM}.

A striking feature observed in \cref{fig:rates}(a) is that the  total reaction rate calculated using our exact quantum  approach for $\lambda=0.95$, $K_\text{tot}=8.33  \times 10^{-11}$ cm$^3$/s,  is 4.1 times smaller than the universal rate of $K_0^\text{UM}=3.45 \times 10^{-10}$ cm$^3$/s in the $s$-wave regime. 
The reaction rate calculated with $\lambda=1.03$ is 4.7 time larger than that with $\lambda=0.95$.
This indicates that non-universal effects due to short-range reflection of the incident scattering flux \cite{Idziaszek:10,Hermsmeier:21,Son:22}   (which appear because not all short-range collisions result in irreversible loss due to, e.g., long-lived complex formation)
can play a crucial role in the ultracold chemical reaction Li~+~NaLi($a^3\Sigma^+$) $\to$ Li$_2$($a^3\Sigma^+_u$)~+~Na despite the presence of a deep potential well.
%, and a large number of open reaction channels.
This stands in contrast to the behavior observed thus far in quantum scattering calculations on other  ultracold atom-molecule insertion reactions \cite{Hutson:07,Croft:17,Makrides:15,Kendrick:21,Kendrick:21b}. In particular, 
% the reactions of  nonpolar \ alkali-dimers such as Li~+~Li$_2$ $\to$ Li$_2$~+~Li populate only a few final product channels, and show the expected non-universal behavior \cite{Hutson:07}. On the other hand,
the chemical reactions of ground-state reactants such as  K~+~KRb($X^1\Sigma^+$) $\to$ K$_2$($X^1\Sigma_g^+$)~+~Rb \cite{Croft:17}, Li~+~LiYb($X^2\Sigma^+$) $\to$ Li$_2$($X^1\Sigma_g^+$) ~+~Yb($^1$S) \cite{Makrides:15}, and Li~+~NaLi($X^1\Sigma^+$) $\to$ Li$_2$($X^1\Sigma_g^+$)~+~Na \cite{Kendrick:21,Kendrick:21b} have deep potential wells and populate a large number of product states, but occur at rates very close (within 30\%) to the universal rates. %those predicted by the UMs.
\Cref{fig:rates}(a) shows that the $\lambda$ dependence of the reaction rate becomes progressively weaker at higher collision energies ($E_\mathrm{c} > 10^{-2}$~K) as quantum interference effects are washed out by multiple partial wave contributions.
%The universal rate predicts well the energy region where the rate begins to increase due to the $p$-wave contribution.
%, implying that the significant $\lambda$-dependent in the s-wave regime related to the zero-energy resonances. }

As shown in \cref{fig:rates}(b),  the chemical reaction Li~+~NaLi($a^3\Sigma^+$) can display both  nearly universal as well as highly non-universal dynamics depending on the value of the scaling parameter $\lambda$ in \cref{pes}. 
The uncertainties in our calculated three-body PES can be estimated at $\pm$5-10\%.
Thus, while non-universal behavior cannot, at present,  be predicted to occur with 100\% certainty, the results  shown in \cref{fig:rates}(b) suggest that its likelihood  is much higher than for the previously studied ultracold chemical reactions, such as Li~+~NaLi($X^1\Sigma^+$) \cite{Kendrick:21}, which did not show any significant deviations from universal behavior as a function of $\lambda$.
Significantly, note that the numerical agreement of the reaction rate with the universal rate does not necessarily mean that the dynamics is well described by the UMs.

\begin{figure}[t]
\begin{center}
\includegraphics[height=0.35\textheight,keepaspectratio]{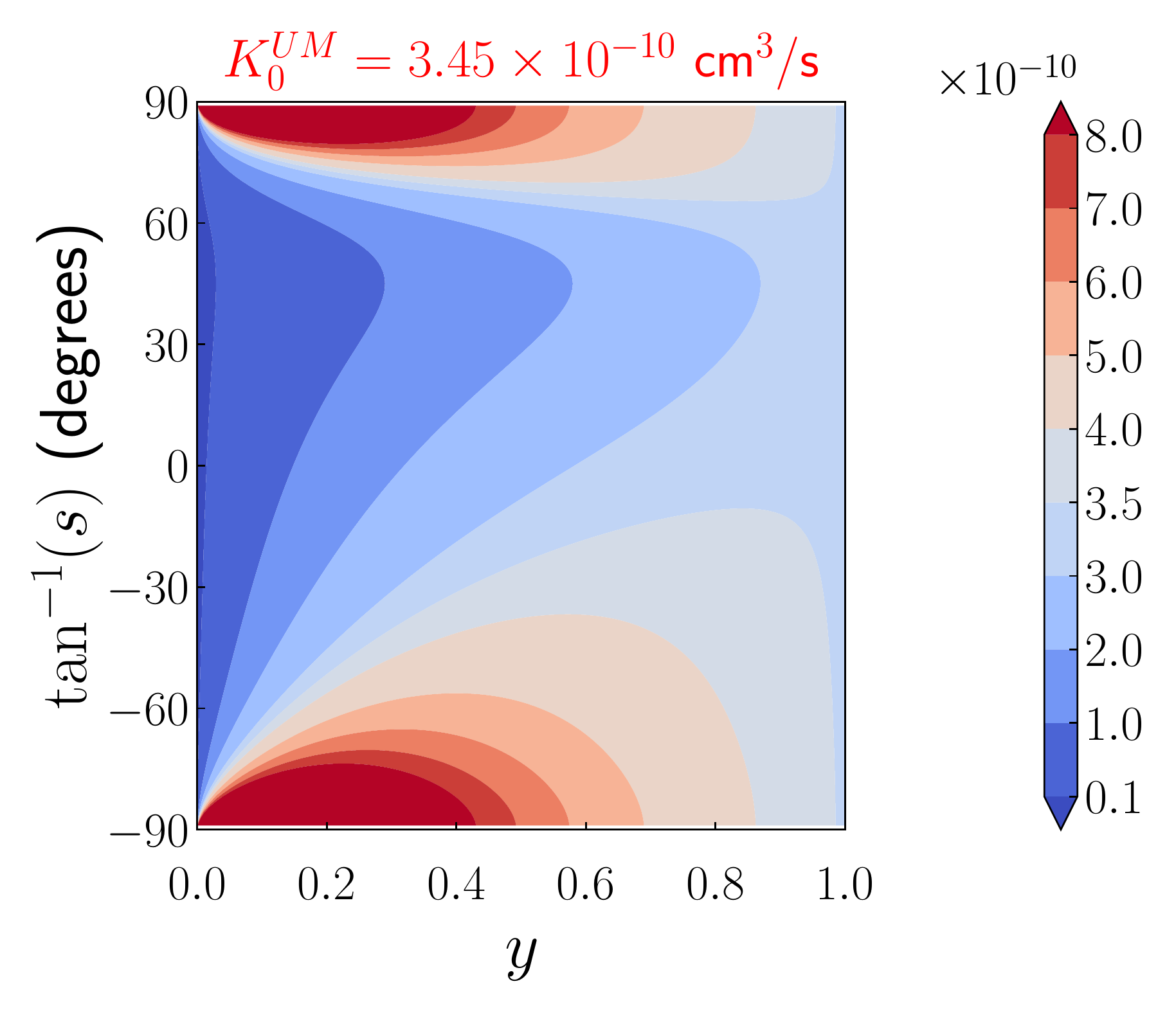}
\end{center}
\caption{
Contour plot of the $s$-wave loss rate (cm$^3$/s) calculated using the UM \cite{Idziaszek:10} as a function of the short-range loss parameter $y\,(0 \leq y \leq 1)$ and the reduced scattering length $s=a/\bar{a}$. 
In the limit $y\to1$, the rate become equal to the universal rate ($K_0^\text{UM}=3.45 \times 10^{-10}$ cm$^3$/s) regardless of the value of $s$. The loss rate approaches zero as $y\to0$.
}
\label{fig:UM}
\end{figure}

To further explore the variation of the rates with the short-range potential, we  plot in \cref{fig:UM} the 
loss rate at the low energy limit using the UM of Ref.~\citen{Idziaszek:10} 
as a function of the short-range loss parameter $y$ and the reduced scattering length $s=a/\bar{a}$, where $\bar{a}$ is the mean scattering length given by\cite{Gribakin_93} $\bar{a}= (2\pi/\Gamma(1/4)^2)({2\mu C_6}/{\hbar^2})^{\frac{1}{4}}$.
% and the range of the scattering length is $-\infty < a < +\infty$. 
We observe that for strong loss ($y>0.8$) it is more likely to observe the rates slightly 
{or below the universal value, as done} in this work. 
In this regime, the small amount of  incident flux reflected from the short range  without reaction
tends to interfere mostly destructively with the incoming flux. The underlying physics is contained in the pole structure of the complex scattering length (see Eq.~(11) of Ref.~\citen{Idziaszek:10}).  On the other hand, for $y<0.5$, observing the rates above the universal value become more and more likely, especially in the case of large  $|a|$ (i.e., near a scattering resonance). This is because  the amount of reflected flux increases  with decreasing $y$, allowing for   a more pronounced quantum interference between the incident and reflected fluxes. As pointed out  by Idziaszek and Julienne \cite{Idziaszek:10}, the quantum interference effects in the UM are counterintuitive, and will be the subject of further study.
% not been studied to the best of our knowledge.
%As pointed out in the original paper by Idziaszek and Julienne, the quantum interference effects in the UM are counterintuitive,  and have not been studied to the best of our knowledge.
 %A detailed analysis of these effects is beyond the scope of this work, and is left for  a future publication.  
 
To explain the strong variation of the ultracold Li~+~NaLi($a^3\Sigma^+$)  reaction rate with $\lambda$ shown in \cref{fig:rates},  we recall (see above)  that both the reactants and products of this chemical reaction are weakly bound vdW molecules, whose binding energies do not exceed 300~K \cite{Gronowski:20}.
By contrast, the ultracold chemical reactions between alkali-metal and alkaline-earth atoms and molecules studied thus far occur between covalently bound molecules, with binding energies above 1000~K.  As a result, the number of open reaction channels populated in these latter reactions (126 for  Li~+~NaLi($^1X$), 143 for K~+~KRb, and 1279 for Li~+~LiYb) \cite{Croft:17,Li:19b,Kendrick:21}.
 %{\it [ Li+NaLi: 122 ch, K+KRb: 140 ch, LiYb+Li: around 600 ch]  } 
is large compared to the 18 open channels in the Li~+~NaLi($a^3\Sigma^+$)  chemical reaction studied here.
%{[\it The same statements are applicable to the inverse process (endothermic reaction with no open channel at E$_c=0$K), thus it would be better to remove "As a results". ]}. 

The total reaction rate is a sum of the individual state-to-state reaction rates, which fluctuate strongly as a function of $\lambda$. 
%due to scattering resonances \cite{Morita:19b,Kendrick:21}. 
These state-to-state fluctuations  average out in the total reaction rate, and the degree of averaging depends on the number of contributing state-to-state rates, which is equal to the number of open reaction channels. Because the number of open channels is 
much smaller for the ultracold Li~+~NaLi($a^3\Sigma^+$)  vdW reaction, 
the averaging is not complete, and the total rate fluctuates strongly as a function of $\lambda$ (see Figs.~S3 and S4 of the SI for the details of state-to-state rates). 
In contrast, the reactions of deeply bound reactants typically populate hundreds of final channels, 
 ensuring nearly complete averaging of the individual state-to-state contributions to the total reaction rate, which then fluctuates only weakly with $\lambda$ (see, e.g., Fig.~12 of Ref.~\citen{Kendrick:21}).
 
This suggests that experimental measurements of state-to-state reaction rates could be used to obtain insight into three-body interactions in triatomic reaction complexes such as  NaLi$_2$. In particular, if a sufficiently large set of experimental state-to-state rates becomes   available, it may be possible to  constrain the form of the three-body interaction (the value of $\lambda$) uniquely. Such state-to-state measurements have recently been carried out  for the  ultracold KRb~+~KRb $\to$ K$_2$~+~Rb$_2$ chemical reaction \cite{Liu:21}. 

%We note that  the $\lambda$-averaged  total reaction rate ($2.15 \times 10^{-10}$ cm$^3$/s) is close to the universal rate 
%{($3.45 \times 10^{-10}$ cm$^3$/s)} as expected  because (thermal)  averaging over many narrow resonances is known to  lead to statistical behavior \cite{Miller:70,Clary:90,Rackham:01,Rackham:03,Alexander:04,Tscherbul:20}, which is manifested in the total reaction probability being a product of the entrance and exit-channel capture probabilities.  The statistical behavior is closely related to the universal behavior considered here (in fact, they are the same in the case of a single open reaction channel, but differ in the multichannel case, where no rigorous definition of the universal behavior is available).

The near-threshold density of states of the {Li-NaLi($a^3\Sigma^+$) trimer} 
can be approximated as $\rho\propto D_e^{3/2}/\sqrt{k_rk_R}$, where $D_e$ is the dissociation energy  %of the trimer PES, 
{measured from the minimum of the trimer PES},
and $k_r$ and $k_R$ are the harmonic force constants for the %atom-diatom and diatom vibrational modes 
NaLi vibrational mode and for the motion along the atom-molecule coordinate $R$
\cite{Frye:21}.
For the Li+NaLi ($^4A'$) reaction complex, we estimate the density of states as $\rho=3.53$ K$^{-1}$, which is 16.7 times lower than that of the  $^{40}$K+$^{40}$K$^{87}$Rb trimer ($\rho=56.6$ K$^{-1}$) \cite{Frye:21} (this calculation neglects the hyperfine structure of the reactants).
%The low density of states % of the Li$_2$Na($^4A'$) trimer 
%implies that the Li~+~NaLi($a^3\Sigma^+$) chemical reaction occurs outside of the highly resonant regime \cite{Mayle:12}.  

This is consistent with our calculations (see \cref{fig:rates}(b) and Fig.~S.3(a) of the SI), which show that state-to-state reaction rates as a function of $\lambda$  are strongly affected by  resonance-like variations. This is because the 5-10\% scaling of the three-body term  leads to a few hundreds of Kelvin changes in the depth of the PES. It is likely that scattering resonances due to near-threshold bound states are responsible for the sensitivity, in which case spectroscopic experiments  should provide important information on the true value of the three-body interaction.
Because the total reaction rate  is the sum of 18 independent state-to-state contributions, we expect that it will show variations on much finer  $\lambda$ scales than shown in \cref{fig:rates} (b).
%Indeed, we observe the rapid and significant variations of the state-to-state as a function of $\lambda$ (see Fig.S3(b) and Fig.S4 in SI), 
 While the $\lambda$ grid interval used in this work is not fine enough to resolve the individual state-to-state oscillations, 
the potential role of near-threshold resonances in causing these oscillations is an important question to be addressed in future work. In addition, it will be important to explore the sensitivity of  state-to-state reaction rates to  the anisotropy of the three-body interaction.  Finally, the background values of reaction rates could be estimated by averaging the calculated $\lambda$-dependent rates over $\lambda$. \cite{Morita:19b}.
%Furthermore, we need to address each state-to-state rate to analyze the resonances especially for precisely separating the background and resonance contributions.

%While the resonances appear with changing the depth of the well of PES, 
%We believe that the sensitivity to the anisotropy will be high regardless of the importance of resonances because the the interaction anisotropy change couplings between the reaction channels at short range. 

\begin{figure}[t]
\begin{center}
\includegraphics[height=0.35\textheight,keepaspectratio]{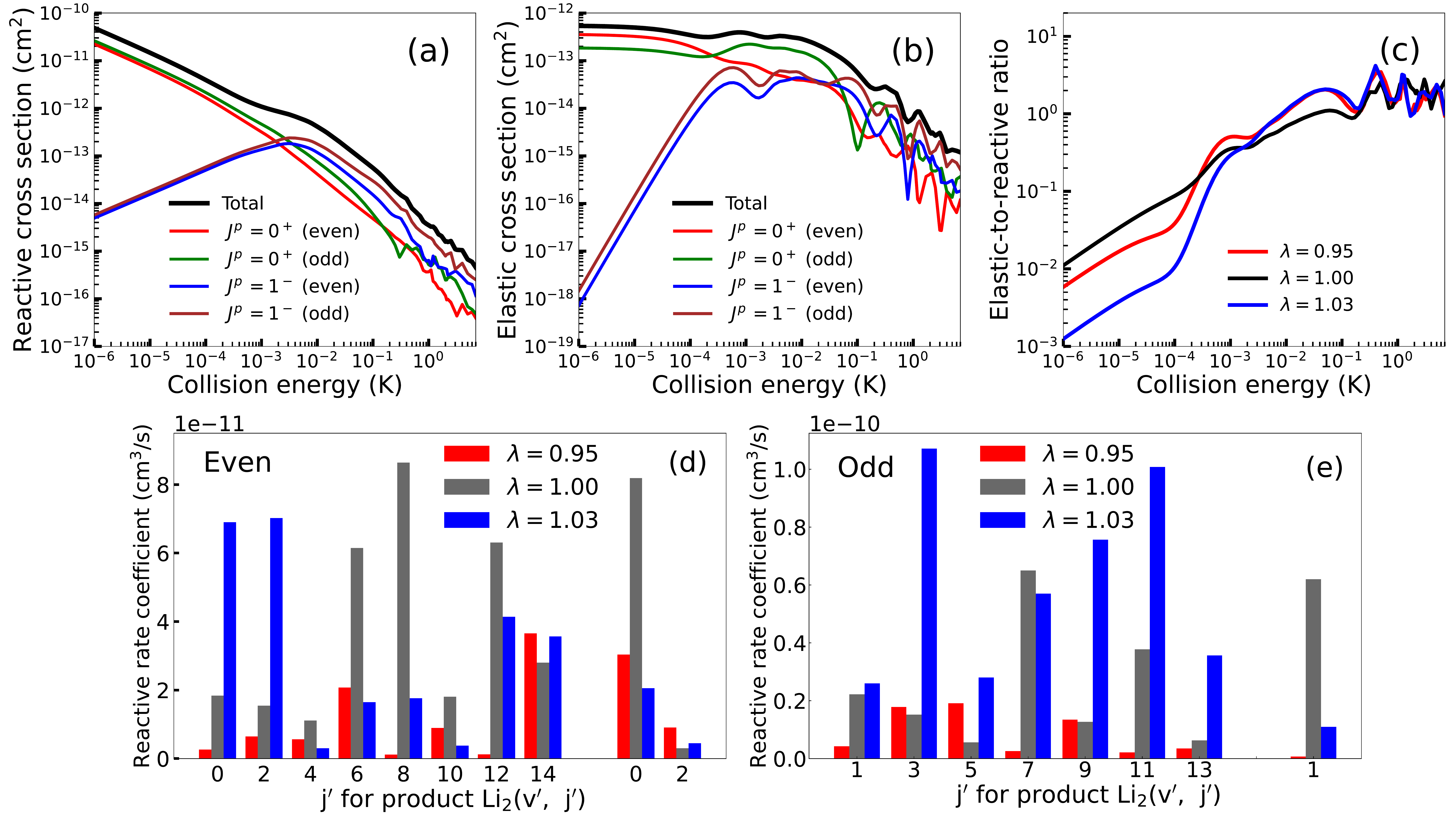}
\end{center}
\caption{
Reactive (a) and elastic (b) cross sections for the  ultracold chemical reaction Li~+~NaLi($v=0,j=0$) $\to$ Li$_2$~+~Na plotted as a function of collision energy for different values of $J$, parity $p$ and exchange symmetry for $\lambda=1$.  The total integral cross sections (solid black lines) are obtained by summing the cross sections for even and odd $^6$Li exchange symmetries multiplied by the statistical factors 1/3 and 2/3.
These factors are the opposite to those used in previous calculations  on the ground-state PESs \cite{Makrides:15,Kendrick:21,Kendrick:21b} because  the  $X^1\Sigma^+_g$ and $a^3\Sigma^+_u$ electronic states of $^6$Li$_2$ are even and odd with respect  to the exchange of identical Li nuclei.
%In the present case, we study Li$_2$($a^3\Sigma^+_u$), thus electronic part of the wave function shows the odd symmetry under exchange of the two identical Li nuclei. 
%On the other hand,  the electronic ground state of $X^1\Sigma^+_g$ studied in the previous papers shows even symmetry under the exchange of the nuclei. 
(c) The ratio of elastic to reactive cross sections $\gamma$ as a function of collision energy for different values of the PES scaling parameter $\lambda$.
Nascent product state distributions of Li$_2$ over final rovibrational states ($v',j'$) at  $E=1$ $\mu$K 
{ for even (d) and odd (e) $j'$  and three values of $\lambda$ indicated in the legend. In (d) and (e), the statistical factors for the exchange symmetries are not used.}
%The contributions of the different $J$, $p$ and exchange symmetry components are color-coded in the same way as in panels (a), and (b).
%Red and green (blue and brown) are even and odd exchange symmetry contributions for $J=0^+$ ($J=1^-$). }
}
\label{fig:XS}
\end{figure}

In \cref{fig:XS}(a) we plot  the total integral cross section for the Li~+~NaLi($a^3\Sigma^+,v=0,j=0$) reaction ($\sigma_r$) as a function of collision energy for several lowest  values of $J$, inversion parity ($p=\pm 1$), and identical Li nuclei exchange symmetry (even/odd).  Both even and odd exchange symmetries are seen to contribute to the reactive cross section nearly equally in the $s$-wave regime ($l=0$), and the $p$-wave contributions ($l=1$) become important at $E>3$~mK. This behavior is consistent with the Wigner threshold scaling  $\sigma_r\simeq E^{l-1/2}$, where $l$ is the orbital angular momentum in the incident channel.
The contribution of the even exchange symmetry to the elastic cross section shown in \Cref{fig:XS}(b) is significantly larger than that of the odd symmetry in the $s$-wave limit, indicating the importance of identical particle permutation symmetry in this reaction.
Quantum interference between the Li exchange and non-exchange
processes is likely responsible for this behavior as observed in H~+~HD $\to$ H~+ HD \cite{Croft:17c,Kendrick:15,SM} 
{(see also the SI \cite{SM})}.
Above 3~mK, the $p$-wave ($l=1$) elastic and reactive cross sections display several broad scattering resonances, whose contribution to the total reaction cross section is largely washed out, 
making it a smooth function of collision energy all the way up to 100 mK.

The ratio $\gamma$ of the elastic to reactive cross sections  displayed in \cref{fig:XS}(c) decreases monotonically in the $s$-wave limit with decreasing collision energy due to the different threshold scaling of the elastic and inelastic cross sections (see above) regardless of the value of $\lambda$. 
A large elastic-to-inelasic collision ratio ($\gamma\geq 10-100$) is an important prerequisite for sympathetic cooling, an experimental technique that relies on momentum-transfer (elastic) collisions of precooled molecules with ultracold atoms to further cool the molecules \cite{Lara:06,Tscherbul:11,Morita:17,Morita:18}. Inelastic collisions and chemical reactions cause trap loss and heating, thereby limiting the efficiency of sympathetic cooling \cite{Lara:06,Tscherbul:11,Morita:17,Morita:18}. 
The results shown in \cref{fig:XS}(c) thus suggest that the chemical reaction Li~+~NaLi($v=0,j=0$) $\to$ Li$_2$~+~Na will prevent efficient sympathetic cooling of NaLi molecules via elastic collisions with spin-polarized Li atoms.
%\cite{Lara:06,Tscherbul:11,Morita:17,Morita:18}. 
%MOVE2 CONCLUSIONS? Thus, vdW chemical reactions of spin-polarized vdW molecules with alkali-metal atoms should be avoided for efficient sympathetic cooling. 
Its worth noting that the efficient sympathetic cooling  observed by Son {\it et al.}~\cite{Son:20} in a trapped mixture of NaLi molecules with Na atoms would not be adversely affected by the  chemical reaction Na~+~NaLi($a^3\Sigma^+,v=0,j=0$) $\to$ Na$_2$($a^3\Sigma^+_u$)~+~Li because it is  endothermic, and hence cannot occur at ultralow temperatures \cite{Hermsmeier:21}. {\it Our results thus suggest that, {in order to achieve efficient sympathetic cooling, the spin-conserving chemical reactions allowed by the Wigner spin rule should  be made  endothermic via, e.g.,} a proper choice of atomic collision partners.}

\Cref{fig:XS}(d) and (e) show nascent Li$_2$ product state distributions over the final rovibrational states ($v^\prime, j^\prime$) at $E=1$~$\mu$K for even and odd exchange symmetries. 
The distributions are highly non-uniform for all $\lambda$ values and  exchange symmetries. 
The state-to-state reaction rates are highly sensitive to $\lambda$ due to quantum interference and resonance effects, which, however, are partially averaged out in the total reaction rate (see above

In summary, motivated by recent  experimental advances in measuring  collisional properties of ultracold NaLi($a^3\Sigma^+$) molecules with alkali-metal atoms  \cite{Son:20,Son:22},
we have performed  accurate  {\it ab initio} and quantum scattering calculations on the prototypical vdW chemical reaction NaLi($a^3\Sigma^+$)~+~Li $\to$ Li$_2$($a^3\Sigma_u^+$)~+~Na, which is energetically allowed at ultralow temperatures. 
We found that the calculated total reaction rate is highly  sensitive to tiny changes in the interaction PES (by a factor of $\simeq 4$), which suggests marked  deviations  from  universal behavior.

These results are unprecedented for an ultracold chemical reaction involving a polar molecule. Indeed, all quantum scattering calculations performed thus far \cite{Ospelkaus:10,Croft:17,Li:19b,Kendrick:21} showed that such reactions do not deviate by more than 30\% from the universal behavior (see, e.g., Fig.~4 of Ref.~\citen{Li:19b} and Fig.~7 of Ref.~\citen{Kendrick:21}).
%In particular, as mentioned above, the exact quantum rate of the K~+~KRb chemical reaction is in perfect agreement with the UM prediction and with experiment \cite{Ospelkaus:10,Croft:17}, and exact quantum rate coefficients of the  chemical reactions Li~+~YbLi($X^2\Sigma^+$) $\to$ Li$_2$($X^1\Sigma_g^+$)~+~Yb and Li~+~NaLi($X^1\Sigma^+$) $\to$ Li$_2(X^1\Sigma^+_g)$~+~Na do not deviate from the universal values by more than 30\% (see Fig.~4 of Ref.~\cite{Li:19b} and Fig.~7 of Ref. \cite{Kendrick:21}).
Therefore, the chemical reaction Li~+~NaLi($a^3\Sigma^+$)  $\to$ Li$_2$~+~Na could be a good candidate for the experimental study of non-universal chemistry in an optically trapped mixture of NaLi($a^3\Sigma^+$) molecules and Li atoms.

A key result of this work is the first observation of {strongly} non-universal dynamics in an ultracold barrierless insertion reaction involving a polar molecule.
The breakdown of universality  is remarkable for several reasons. First, it gives us a rare  glimpse into a new {regime} of ultracold chemical dynamics, which cannot be described by simple universal models even at a qualitative level.
Second, non-universal dynamics are sensitive to fine details of the underlying PES of the reaction complex. As such, measuring state-resolved reaction rates beyond the universal limit, as done in recent pioneering experiments \cite{Son:22,Park:23} will enable high-precision characterization of intermolecular interactions within metastable reaction complexes, a much sought-after goal of ultracold chemistry \cite{Krems:08,Balakrishnan:16,Bohn:17}.

Our results also bear significant implications for molecular sympathetic cooling of alkali-dimer molecules by collisions with ultracold atoms in a magnetic trap. The large reaction rates of spin-polarized reactants observed here (see \Cref{fig:XS}(c)) imply that preparing the reactants in their fully spin-polarized initial states will not prevent rapid collisional losses if spin-conserving atom-molecule chemical reactions (such as the Li~+~NaLi($a^3\Sigma^+$) reaction explored here)  are energetically allowed.  This leads to a general design principle for choosing ultracold atom-molecule systems suitable for sympathetic cooling experiments:  Avoid  spin-conserving chemical reactions by  choosing  the  coolant atom in such a way as to make the reaction endothermic, and hence energetically forbidden at ultralow temperatures.

\section*{Supporting Information}
See the supporting information associated with this article for details of {\it ab initio} calculations of the $^4A'$ PES of the Li$_2$Na trimer and for technical details of quantum reactive scattering calculations.

%%%=============================================================================
\begin{acknowledgement}

M.M. thanks Dr. Pablo G. Jambrina for useful discussions.
 This work was supported by the NSF through the
CAREER program (PHY-2045681) and by  the U.S. Air Force Office
for Scientific Research (AFOSR) under Contracts Nos. FA9550-19-1-0312 and FA9550-22-1-0361 to P.B. and T.V.T.  J.K. and S.K. acknowledge support from the U.S. AFOSR under Grant No. FA9550-21-1-0153 and the NSF under Grant No. PHY-1908634. 
B.K.K. acknowledges that part of this work was done under the auspices
of the US Department of Energy under Project No. 20170221ER
of the Laboratory Directed Research and Development Program at Los
Alamos National Laboratory.  This work used resources provided by the Los Alamos National Laboratory Institutional Computing Program. 
Los Alamos National Laboratory is operated
by Triad National Security, LLC, for the National Nuclear Security
Administration of the U.S. Department of Energy (contract No.
89233218CNA000001).

\end{acknowledgement}

\section*{Data availability statement}
The data that support the findings of this study are available within the article and its
supporting information. Data is also available from the authors upon reasonable request.

%%%%%%%%%%%%%%%%%%%%%%%%%%%%%%%%%%%%%%%%%%%%%
\bibliography{cold_mol_new}

\begin{mcitethebibliography}{78}
\bibitem[Not()]{Note-1}
A notable exception is vibrational excitation of the reactant molecule, for
  which abstraction reactions can also become barrierless and exhibit a
  potential well (e.g., H(D) + H$_2$($v\geq5$)) (See
  Refs.~\citen{Simbotin:15,Kendrick:18,Kendrick:19}).\relax
\mciteBstWouldAddEndPunctfalse
\mciteSetBstMidEndSepPunct{\mcitedefaultmidpunct}
{}{\mcitedefaultseppunct}\relax
\EndOfBibitem
\end{mcitethebibliography}
\end{document}

%\begin{tocentry}
%\vspace{-50pt}
%\includegraphics[scale=0.4, trim = 130 0 0 140 ]{Fig_TOC.pdf}. %temporary
%\includegraphics[scale=1.5,  trim = 30 0 0 0]{Fig_TOC.pdf}. %temporary
%\includegraphics[scale=0.3]{Fig_TOC.pdf}. %temporary
%\end{tocentry}

\singlespacing
In this Supporting Information (SI), we present the details of {\it ab initio} calculations for the potential energy surface (PES) of the NaLi$_2$ (1$^4$A$^{\prime}$) reaction complex (\cref{sec:SI_PES}). 
\cref{sec:SI_CC,sec:SI_computation} provide the methodological and computational details of our quantum reactive scattering calculations using the APH3D code. 
\cref{sec:SI_computation} discusses convergence tests.
\cref{sec:SI_UM} describes the universal loss (total reaction) rate, along with the values of parameters for the Li+NaLi($a^3\Sigma^+$) system. 
The $\lambda$ dependence of the reaction rate (including the effects of exchange symmetry of the Li nuclei) is discussed in detail in \cref{sec:SI_scaling}, which also contains the state-to-state reaction rates. 
The reactive and elastic cross sections and elastic-to-inelastic ratios calculated with the pairwise PES ($\lambda=0$) are presented in \cref{sec:results_2body}.

%on the ultracold chemical reaction Li($^2$S)~+~NaLi($a^3\Sigma^+$) $\to$ Li$_2$($a^3\Sigma_u^+$)~+~Na($^2$S).
% and describes the convergence tests.

%Section I is devoted to the ab initio calculations whereas Sec. II to quantum scattering calculations.
%\renewcommand{\thefigure}{\Alph{section}-\arabic{figure}} 

%\setcounter{equation}{0}
%\setcounter{figure}{0}

\begin{center}
\textbf{\large
\section{\label{sec:SI_PES} Potential energy surface}
}
\end{center}

The spin conserving reaction Li($^2$S) + NaLi($a^3\Sigma^+,v=0,\,j=0$) $\to$ Li$_2$($a^3\Sigma_u^+,v',\,j'$)~+~Na($^2$S) proceeds on the $1 ^4A^\prime$  PES. 
The calculations of the  PES, $V_{1^4A^{\prime}}(r_a,r_b,\alpha)$, have been performed on a set of discrete grids of NaLi$_2$ geometries described by two NaLi bond lengths $r_a$ and $r_b$ and the bond angle $\alpha$ for Li($a$)-Na-Li($b$).  
The grid consists of bond lengths $r_a$ and $r_b$ from $3.75 a_0$ to $14 a_0$ with a step of $0.25 a_0$ (where $a_0$ is the Bohr radius), and the bond angle $\alpha$ from $0^\circ$ to $180^\circ$ with a step of $5^\circ$ with additional $\alpha$ values of $1^\circ$, $2^\circ$, $4^\circ$, $8^\circ$ and $179^\circ$ for near collinear geometries.  We performed  the  electronic structure calculations using MOLPRO~\cite{MOLPRO}.

In our reactive scattering calculations the PES is represented as the sum of pairwise (two-body) and three-body interactions
\begin{equation}
{
V_{1 ^4A^\prime}(r_\textrm{Li-Li$^\prime$},r_\textrm{Na-Li},r_\textrm{Na-Li$^\prime$})={V^\textrm{pairwise}_{1 ^4A^\prime}(r_\textrm{Li-Li$^\prime$},r_\textrm{Na-Li},r_\textrm{Na-Li$^\prime$})}+\lambda {V^\textrm{3-body}_{1 ^4A^\prime}(r_\textrm{Li-Li$^\prime$},r_\textrm{Na-Li},r_\textrm{Na-Li$^\prime$})}
},
\label{eq:Vfull}
\end{equation}  
where arguments specifying the geometry of NaLi$_2$ are transformed into the three bond lengths to match the interface of the APH3D code (see \cref{sec:SI_CC}), $r_\textrm{A-B}$ denotes the bond distance between A and B, the prime for Li$^\prime$ is introduced to make a distinction of two identical Li nuclei. 
The pairwise (two-body) interaction $V^\textrm{pairwise}_{1 ^4A^\prime}$ is decomposed into the sum of diatomic interaction potentials as
\begin{equation}
{
V^\textrm{pairwise}_{1 ^4A^\prime} (r_\textrm{Li-Li$^\prime$},r_\textrm{Na-Li},r_\textrm{Na-Li$^\prime$})=V^\textrm{Li$_2$}_{a ^3\Sigma_u^+}(r_\textrm{Li-Li$^\prime$})+V^\textrm{NaLi}_{a ^3\Sigma^+}(r_\textrm{Na-Li})+V^\textrm{NaLi}_{a ^3\Sigma^+}(r_\textrm{Na-Li$^\prime$}),
}
\label{eq:2body}
\end{equation}  
where $V^\textrm{A-B}$ in the right-hand-side of the equation denotes the potential energy curve for the diatomic AB molecule. 
This additive pairwise potential $V^\textrm{pairwise}_{1 ^4A^\prime}$ dominates the long-range behaviour of the total potential of $V_{1 ^4A^\prime}$. 
For $V^\textrm{NaLi}_{a ^3\Sigma^+}$ and $V^\textrm{Li$_2$}_{a ^3\Sigma_u^+}$ in \cref{eq:2body}, we employ the accurate {\it ab initio} potential energy curves \cite{Lesiuk:20,Hermsmeier:21}.
The non-additive three-body interaction, $V^\textrm{3-body}_{1 ^4A^\prime}$, governs the short-range dynamics. We use accurate electronic structure calculations to obtain the three-body interaction. The calculations and the fitting procedure are described below.
Because $V^\mathrm{pairwise}$ and $V^\mathrm{3-body}$ are smooth functions of the internal coordinates, so is the total potential $V$ given in \cref{eq:Vfull} at any $\lambda$.

The details of the present electronic structure calculations are the same as previously published for the spin doublet NaLi$_2$ potentials~\cite{Kendrick:21}. During the calculations of the global $1^4A'$ potential we encountered a conical intersection and an avoided crossing between the $1^4A'$  and $2^4A'$ states located on the repulsive wall of the PES, but still at relatively low energies to affect the fit. The appearance of the conical intersection causes difficulties in the fitting procedure of the three-body term by expansion in Legendre polynomials and interpolation of the 2-dimensional radial coefficients as functions of bond distances. Because we fit the lowest $1^4A'$ adiabatic PES instead of the diabatic one, we need a procedure to smooth out numerical instabilities caused by abrupt changes of the potential in the vicinity of the conical intersection seam that project also to long distances. To smooth out the kinks in the RKHS fits of the three-body Legendre expansion \cite{Kendrick:21} we resorted to smoothly connecting the analytical Axilrod-Teller three-body dispersion term \cite{Pena:80,Axilrod:43} to the short-range three-body fit. In order to accomplish this, we need the $C_9$ long-range dispersion coefficient for the NaLi$_2$ ($1^4A^{\prime}$) trimer.
% calculated as described below.

We calculated the $C_9$ long-range dipole dispersion coefficient of the $V^\textrm{3-body}_{1 ^4A^\prime}$ potential using imaginary frequency-dependent dipole polarizabilites of Li and Na atoms of Derevianko {\em et al.}~\cite{Derevianko:2010} using the following formula:
\begin{equation}
C_9=\frac{3}{\pi}\int\alpha_{\mathrm{Li}}(i\omega)\alpha_{\mathrm{Li}}(i\omega)\alpha_{\mathrm{Na}}(i\omega) d\omega.
\label{eq:C9}
\end{equation}

We used 50 Gauss-Legendre quadrature nodes for the numerical evaluation of the above integral and we obtained the value of $C_9=175597.73$ $E_ha_0^9$, which is close to the literature value of Pe{\~n}a {\em et al.}~\cite{Pena:80} of $17.566\times10^4$ $E_ha_0^9$.  
This procedure has a minimal effect on the global minimum region of the full  $1^4A'$ potential.

To obtain an analytical representation of  $V_{1^4A^{\prime}}(r_a,r_b,\alpha)$ from the {\it ab initio} grid points, we apply a fit or interpolation procedure similar to that described in Ref.~\cite{Kendrick:21}. 
Namely, we first obtain the non-additive three-body contribution by subtracting the pairwise (two-body) potential constructed from the diatomic NaLi($a^3\Sigma^+$) and Li$_2$ ($a^3\Sigma_u^+$) potentials also obtained by MOLPRO calculations.
This non-additive three-body contribution is then expanded in a series of Legendre polynomials $P_l(\cos\alpha)$ with $l$ up to $l_{max}=8$, where the two-dimensional $v_l(r_a,r_b)$ radial expansion coefficients are interpolated.
The short-range three-body contribution is smoothly connected using a switching function to the long-range Axilrod-Teller dispersion potential $V_{{\rm 3b}}^{\rm lr}=C_9(1+\cos\alpha_1\cos\alpha_2\cos\alpha_3)/(r_1^3r_2^3r_3^3)$, where $r_i$ and $\alpha_i$ are the lengths of the sides and  inner angles of the triangle formed by the three atoms and the van der Waals three-body dispersion coefficient $C_9$ is computed from the imaginary-frequency dependent polarizabilities~\cite{Derevianko:2010} of the Na and Li atoms as described above.
Here, $E_{\rm h}$ is the Hartree energy. The switching function is the sigmoid function $sw(r_i)=(1-\tanh[\beta_{\rm sw}(r_i-r_{\rm sw})])/2$ 
for $i=1$, 2, or 3 where $r_{\rm sw}=10.0\,a_0$ and $\beta_{\rm sw}=1.0/a_0$ for all $i$.

\begin{center}
\textbf{ \large
\section{\label{sec:SI_CC} Quantum reactive scattering calculations}
}
\end{center}

%RESUME HERE

The quantum mechanical dynamics of the Li($^2$S) + NaLi($a^3\Sigma^+,v=0,\,j=0$) $\to$ Li$_2$($a^3\Sigma_u^+,v',\,j'$)~+~Na($^2$S) reaction on the $1 ^4A^\prime$ PES is explored by performing the rigorous approach based on the adiabatically adjusting principal axes hyperspherical (APH) coordinates method \cite{Pack:87, Kendrick:99}. We only briefly summarize the method and calculation procedures since several publications describe the detailed information about the APH coordinate method and its implementation (APH3D program suite) \cite{Kendrick:2018_Nonad, Makrides:15, Kendrick:21}. 
The values of parameters used in the present scattering calculations are described in \cref{sec:SI_computation}.

The time-independent Schr\"{o}dinger equation for the triatomic system is written in the hyperspherical coordinates as 
\begin{equation}
\left[-\frac{\hbar^2}{2\mu\rho^5}\frac{\partial}{\partial \rho}\rho^5\frac{\partial}{\partial \rho}+\frac{\hat{\Lambda}^2}{2\mu\rho^{2}}+V(\rho,\Tilde{\theta},\phi)\right]\Psi(\rho,\Tilde{\theta},\phi,\alpha,\beta,\gamma) = E \Psi(\rho,\Tilde{\theta},\phi,\alpha,\beta,\gamma),
\label{eq:Schrodinger}
\end{equation}
where $\mu$ is the three-body reduced mass defined by using the mass of each atom as $\sqrt{m_Am_Bm_C/(m_A+m_B+m_C)}$,  $\hat{\Lambda}$ is the grand angular momentum operator, $V$ denotes the PES, $E$ denotes the total energy,
$\rho$ is the hyperradius that is the radial coordinate corresponding to a symmetric stretch motion of the triatomic complex, $\Tilde{\theta}$ is the polar angle corresponding to a bending motion and related to the $\theta$ of Pack and Parker as $\Tilde{\theta}=\pi-2\theta$, $\phi$ is the azimuthal angle corresponding to an internal pseudo-rotational motion and related to the $\chi$ of Pack and Parker as $\phi=2\chi$, the three Euler angles $\alpha, \beta, \gamma$ specify the orientation of the body-fixed (BF) frame relative to the space-fixed (SF) frame. The z-axis of the BF frame directs perpendicular to the plane of triatomic complex and the x and y axes are chosen to lie along the instantaneous principal axes of inertia. 
For a given total angular momentum $J$, inversion parity $p$ and exchange symmetry $q$ for the identical atoms, the wave function $\Psi$ is expanded using the 5D surface functions $\Phi$ that are the functions of hyperangles $\Tilde{\theta}$ and $\phi$, and three Euler angles as
\begin{equation}
\Psi^{JMpq}(\rho,\Tilde{\theta},\phi,\alpha,\beta,\gamma)=4\sqrt{2}\sum_{n}^{N_\mathrm{ch}} \rho^{-5/2} \zeta^{Jpq}_{n}(\rho)\Phi^{JMpq}_{n}(\Tilde{\theta},\phi,\alpha,\beta,\gamma;\rho_{\xi}), 
\label{eq:wavefunction}
\end{equation}
where the $N_\mathrm{ch}$ specifies the number of coupled channels, and $\zeta$ are the $\rho$-dependent expansion coefficients. The surface functions are obtained by solving the following eigenproblem at each $\rho_\xi$,
\begin{equation}
\left[\frac{\Tilde{\Lambda}^2}{2\mu\rho^{2}_{\xi}}+\frac{15\hbar^2}{8\mu\rho^{2}_{\xi}}+V(\rho_{\xi},\Tilde{\theta},\phi)\right]\Phi^{JMpq}_{n}(\Tilde{\theta},\phi,\alpha,\beta,\gamma;\rho_{\xi}) = \epsilon^{Jpq}_{n}(\rho_{\xi}) \Phi^{JMpq}_{n}(\Tilde{\theta},\phi,\alpha,\beta,\gamma;\rho_{\xi}).
\label{eq:SI_surface}
\end{equation}

Outside the strongly interacting short-range region, where the inter-arrangement coupling is negligible, we use the Delves hyperspherical coordinates. 
In contrast to \cref{eq:wavefunction}, the total wave function is expanded in a set of the products of $\rho$-dependent vibrational wavefunctions and the angular functions in each arrangement channel of $\tau$ as
\begin{equation}
\Psi^{JMpq}(\rho,\theta,\hat{S_\tau},\hat{s}_\tau)=2\sum_{n}^{N_\mathrm{ch}} \rho^{-5/2} \zeta^{Jpq}_{n}(\rho) \frac{\Upsilon^{Jq}_n(\theta_\tau;\rho)}{\sin{2\theta_\tau}} \mathcal{Y}^{JMpq}_n(\hat{S_\tau},\hat{s}_\tau), 
\label{eq:FD_wavefunction}
\end{equation}
where $\Upsilon$ denotes vibrational wavefunction, $ \mathcal{Y}$ denotes coupled-angular momentum with the mass-scaled Jacobi vectors $\bm{S}_\tau$ and  $\bm{s}_\tau$ in the arrangement channel of $\tau$. Although the hyperradius $\rho$ is the same as in the APH coordinates, the hyperangles are different and depend on $\tau$ and are defined as $\theta_\tau=\tan^{-1}(s_\tau/S_\tau)$ and $\Theta_\tau=\cos^{-1}(\bm{S_\tau}\cdot\bm{s_\tau}/S_\tau s_\tau)$. We note that the above equation is in the SF coordinate representation. The 1D equation for the vibration is given as
\begin{equation}
\left\{-\frac{\hbar^2}{2\mu\rho^2_\xi}\left[\frac{\partial^2}{\partial\theta^2_\tau}-\frac{j_\tau(j_\tau+1)}{\sin{\theta_\tau}^2}-\frac{l_\tau(l_\tau+1)}{\cos{\theta_\tau}^2}\right]+V_\tau(\rho_\xi \sin{\theta_\tau}) \right\} \Upsilon_n(\theta_\tau;\rho_\xi) = \mathcal{E}_n(\rho_\xi) \Upsilon_n(\theta_\tau;\rho_\xi) , 
\label{eq:FD_vib}
\end{equation}
where $j_\tau$ is rotational quantum number of the diatomic molecule, $l_\tau$  is the orbital angular momentum of the atom around the center of mass for the diatomic molecule,  $V_\tau$ is the two-body potential for the molecule in the arrangement $\tau$. The equation parametrically depends on the values of $j_\tau$ and $l_\tau$, thus the index $n$ is specified with collective quantum numbers of $\{v_\tau, j_\tau, l_\tau \}$, where $v_\tau$ is the quantum number for the vibration.

In both the APH and Delves %(FD)}  
coordinates, the coupled-channel equations to be satisfied by the expansion coefficients of $\zeta(\rho)$ are obtained by substituting the \cref{eq:wavefunction,eq:FD_wavefunction} 
into the total wave function $\Psi$ in the Schr\"{o}dinger equation. The Hamiltonian in the Delves coordinates is similar to that in the APH coordiantes \cref{eq:Schrodinger} but the expression for $\hat{\Lambda}^2$ is different \cite{Parker:02}.
The obtained coupled-channel equations are solved based on the log-derivative propagation method followed by the matching procedure to the scattering boundary condition at a large asymptotic $\rho$ value to obtain the S-matrix \cite{Pack:87,Parker:02}.

\begin{center}
\textbf{ \large
\section{\label{sec:SI_computation}  Computational details}
}
\end{center}

\begin{figure}[b!]
\begin{center}
\includegraphics[height=0.27\textheight,keepaspectratio]{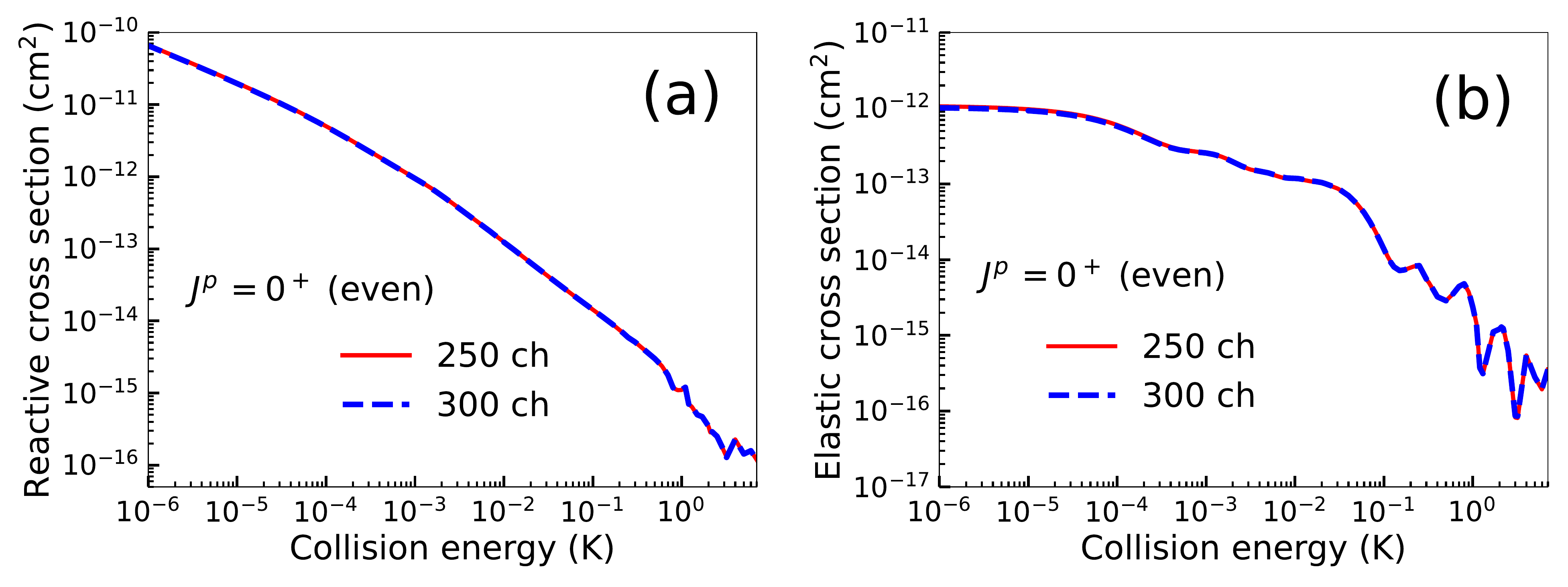}
\end{center}
\caption{
Reactive (a) and elastic (b) cross sections as a function of collision energy calculated using  coupled-channel basis sets containing 250 and 300 channels for $J=0^+$ and even exchange symmetry. The statistical factor of 1/3 is not applied. 
}
\label{figSI:NChannels}
\end{figure}
All quantum mechanical scattering calculations in this study are performed with the APH3D program suite developed at  the Los Alamos National Laboratory. Here, we summarize the relevant parameters employed in our calculations.
The masses for $^{23}$Na and $^6$Li are 22.9897692808 amu and  6.0151227945 amu, respectively. 
At short range ($6 \leq \rho \leq 40 a_0$) the time independent Schr\"{o}dinger equation is described in the APH coordinates, and the range is divided into  0.1 $a_0$ wide sectors. 
The surface functions to expand the total wave function within a sector are obtained by solving the eigenproblem at the center of each sector as described below. The log-derivative propagation is carried out using the diabatic-by-sector method with   200 propagation steps per each sector.

To construct the Hamiltonian matrix for the eigenproblem, the hyperangles are expanded with a primitive basis of Jacobi polynomials in $\tilde{\theta}$ and complex exponential functions in $\phi$ whose sizes are specified by $l_\text{max}=159$ and $m_\text{max}=320$ \cite{Kendrick:99}.
The actual numbers of basis functions are given by $l_\text{max}+1$ and $2m_\text{max}+1$, respectively, and a the discrete variable representation (DVR) is used in $\tilde{\theta}$ rather than the finite basis representation. 
While the original dimension of the Hamiltonian matrix is 102560 for $J=0^+$ (even), actual dimension of the matrix is typically reduced to 40000-60000 by employing the Sequential Diagonalization Truncation (SDT) method. 
The number of channels used in the coupled-channel calculation is 250 with $J=0$ for each inversion and exchange symmetry. We increase the number of channels to 300 for $J=1$.

\Cref{figSI:NChannels} shows the convergence of the reactive and elastic cross sections as a function of collision energy with 250 channels for $J=0^+$ with even exchange symmetry.  
Beyond $\rho=40 a_0$, there is no interaction between the different reaction arrangements (see main text), and the coupled channel equations are solved in the Delves coordinates as described above.
The three-body term decays rapidly with increasing $\rho$ and becomes negligible at $\rho\geq 22 a_0$. Thus, the PES and the associated adiabatic potentials do not depend on $\lambda$ at $\rho\geq 22 a_0$.
The number of channels, propagation sectors, and log-derivative propagation steps within each sector used in the Delves propagation are the same as those used in the APH region.   

\begin{figure}[b!]
\begin{center}
\includegraphics[height=0.27\textheight,keepaspectratio]{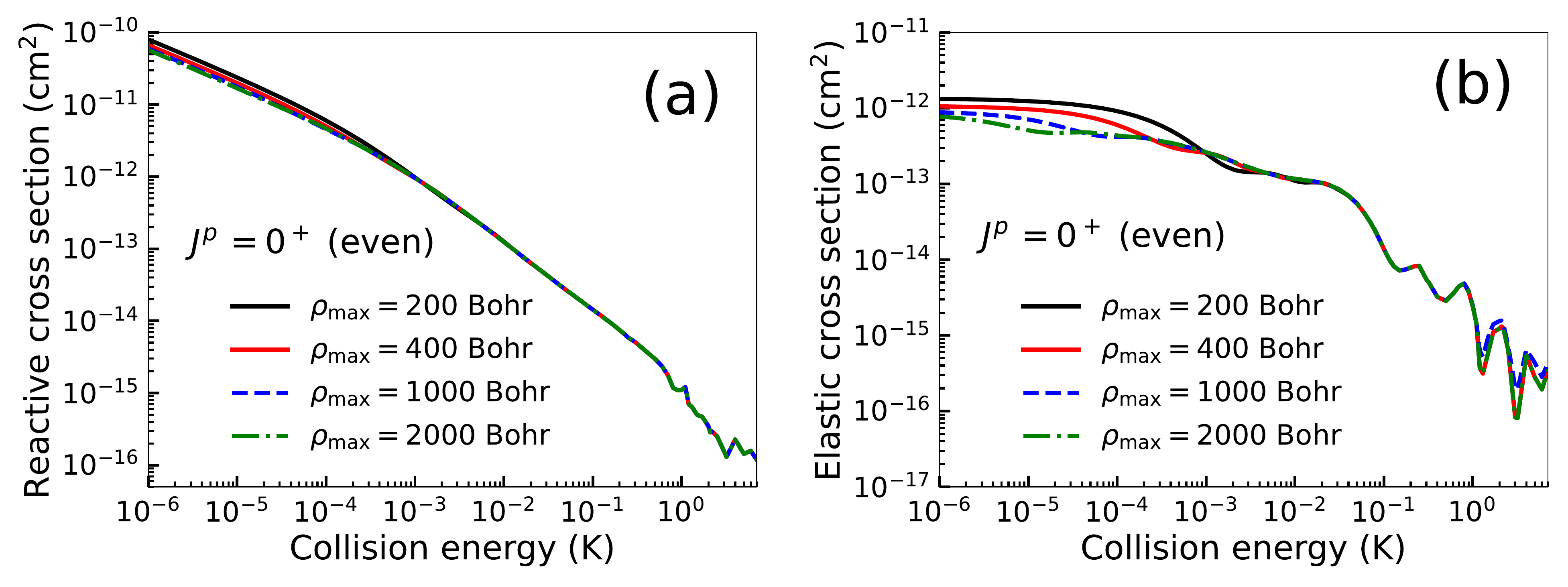}
\end{center}
\caption{
Reactive (a) and elastic (b) cross sections as a function of collision energy for different maximum hyper radius ($\rho_\mathrm{max}$) in the log-derivative propagation for solving the coupled-channel equation for $J=0^+$ and the even exchange symmetry. Statistical factor of 1/3 is not multiplied. 
}
\label{figSI:rhomax}
\end{figure}

After reaching the end of the propagation grid at $\rho_\mathrm{max}=400 a_0$, the reactance (K) matrix is calculated in the asymptotic Jacobi basis, and the full state-to-state S-matrix is obtained from the K-matrix as described in Ref.~\cite{Parker:02}. 
\Cref{figSI:rhomax} shows the $\rho_\mathrm{max}$ dependence of the calculated reactive and elastic cross sections.
Despite the relatively slow $\rho_\mathrm{max}$ convergence in the low energy region observed in \cref{figSI:rhomax}(a), we can still obtain an accurate reactive cross section with $\rho_\mathrm{max}=400 a_0$. On the other hand,  the elastic cross section exhibits extremely slow convergence with respect to $\rho_\mathrm{max}$, especially in the collision energy range between 10$^{-5}$ to 10$^{-4}$ K. 
%This slow convergence  is counterintuitive. 

% and might be due to an unexpected numerical instability or the very diffuse nature of the dimer vibrational states that is unique to this system.

The elastic cross section is typically more difficult to converge, however, the
very large $\rho_\mathrm{max} > 1000 a_0$ is counterintuitive. Small oscillations in the elastic cross sections are seen
in the black, red, blue and green curves near 2~mK, 0.4~mK, 0.1~mK and 0.02~mK, respectively. These oscillations
are most likely due to the tiny difference between the Delves energies computed at $\rho_\mathrm{max}$ and the asymptotic Jacobi energies.
This energy step discontinuity is due to the difference between the two coordinate systems, for large $\rho$ the one-dimensional diatomic potential energy curve for each arrangement channel in Delves coordinates is slightly different than the asymptotic Jacobi one. 
As $\rho_\mathrm{max}$ is increased the energy step decreases, which is consistent with the observed oscillations moving to lower collision energies as $\rho_\mathrm{max}$ is increased. 
The energy step is always present and can lead to unphysical self-interference effects that appear to be more significant for this system.

We conclude that it is necessary to use a vary large $\rho_\text{max}$ to achieve complete numerical  convergence of the elastic cross section  for ultracold vdW reactions. These considerations only apply to the elastic cross sections. As such, they do  not affect the discussions and conclusions described in this main text because our primary interest is in  the reaction processes and order-of-magnitude estimates of the elastic-to-inelastic ratio. 

%The most likely reason for the oscillations in the elastic cross section at $\simeq$10$^{-5}$ Kelvin is the small energy mismatch  between the energies at $\rho_\mathrm{max}$ in the Delves and Jacobi coordinates due to the difference in these coordinate systems. The energy mismatch decreases with increasing $\rho_\mathrm{max}$, but unphysical self-interference in the elastic channel remains at the low collision energies.

% This energy step is always a problem but it has not been reported the emergence of the energy step problem causing the unphysical wiggles and resultant slow convergence with respect to $\rho_\mathrm{max}$ for the elastic cross section. 
%[{Brian, please explain more on the origin of the energy step if necessary. Also please feel free to put additional figures/data in this section to discuss about $\rho_\mathrm{max}$-dependence. But, in this paper, all the results in the main text are obtained with $\rho_\mathrm{max}=400 a_0$}. It would be better to publish complete discussion and analysis after we find the same problem at least in a different system/reaction.]. 

\begin{center}
\textbf{ \large
\section{\label{sec:SI_UM} Universal rate}
}
\end{center}

\begin{equation}
{
K_{l=0}^\text{UM}\, =\, \frac{4\pi\hbar}{\mu} \bar{a}
}
\label{eq:bar_a}
\end{equation}  

\begin{equation}
{
\bar{a}\, =\, \frac{2\pi}{\Gamma(1/4)^2}(\frac{2\mu C_6}{\hbar^2})^{\frac{1}{4}}
}
\label{eq:bar_a}
\end{equation}

Here, we describe the universal rate for the $s$-wave ($l=0$) $K_0^\text{UM}$ and  $p$-wave ($l=1$) $K_1^\text{UM}(E)$ based on the universal model (UM).~\cite{Idziaszek:10,Julienne:11}
The energy independent s-wave universal rate is 
\begin{equation}
{
K_0^\text{UM} = \frac{4\pi\hbar}{\mu} \bar{a},
}
\label{eq:K0}
\end{equation}  
where $\mu$ is the reduced mass for the $^6$Li+$^{23}$Na$^6$Li system, the mean scattering length $\bar{a}$ is given as \cite{Gribakin_93}
\begin{equation}
{
\bar{a}\, =\, \frac{4\pi R_6}{\Gamma(1/4)^2}\, =\, \frac{2\pi}{\Gamma(1/4)^2}(\frac{2\mu C_6}{\hbar^2})^{\frac{1}{4}},
}
\label{eq:bar_a}
\end{equation}  
where $R_6=(2\mu C_6/\hbar^2)^{\frac{1}{4}}/2$ is a characteristic length.
As stated in the main text, we use the {\it ab initio} value $C_6=2891.15$ $E_h/a_0^6$ for the long-rang dispersion coefficient between Li and NaLi.
The universal reaction rate is $K_0^\text{UM}=3.45 \times 10^{-10}$ cm$^3$/s in the $s$-wave regime. 
%Also, the universal rate predicts well the energy dependence in the region where the rate begins to increase with increasing the energy due to 
The $p$-wave contribution to the universal rate is \cite{Idziaszek:10,Julienne:11}
\begin{equation}
{
K_1^\text{UM} (E) = \frac{12\pi\hbar}{\mu} (k\bar{a})^2 \bar{a_1},
}
\label{eq:K1}
\end{equation}  
where $\bar{a_1}=\bar{a}\Gamma(1/4)^6/(144\pi^2\Gamma(3/4)^2)$. The $p$-wave contribution is negligible compared to $K_0$ as $E \to 0\, (k \to 0)$.
Note that the universal rate does not depend on the short-range behavior of the PES.

\begin{figure}[t!]
\begin{center}
\includegraphics[height=0.27\textheight,keepaspectratio]{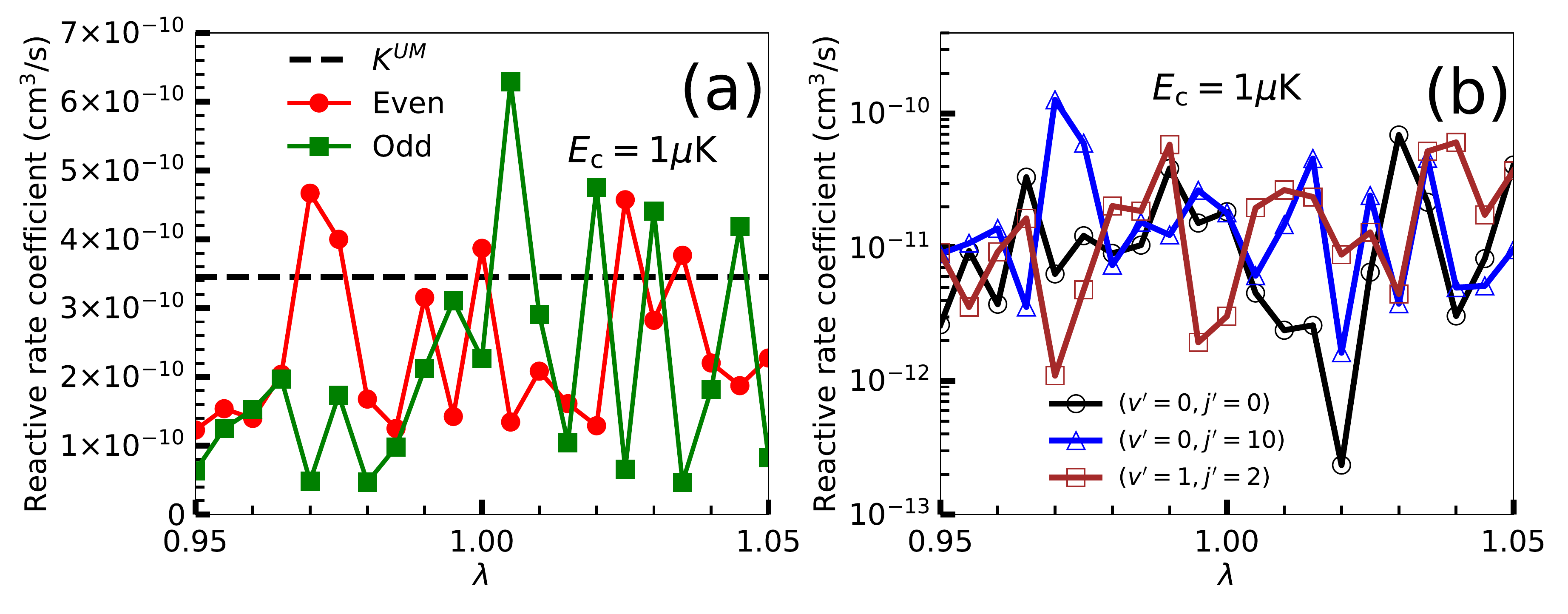}
\end{center}
\caption{
(a) Total reaction rate as a function of the scaling parameter $\lambda$ at $E_\mathrm{c}=10^{-6}$K (1$\mu$K) for even and odd exchange symmetries of identical Li nuclei.
The statistical factors of 1/3 and 2/3 for even and odd symmetries are not applied. The dashed  line indicates the universal rate, $K^\text{UM}=3.45 \times10^{-10}$ cm$^3$/s. 
(b) State-to-state reaction rates to the final states ($v^\prime=0,\,j^\prime=0$), ($v^\prime=0,\,j^\prime=10$), and ($v^\prime=1,\,j^\prime=2$) as a function $\lambda$ at $E_\mathrm{c}=10^{-6}$K for the even exchange symmetry and $J=0^+$. 
}
\label{figSI:scaling}
\end{figure}

\begin{center}
\textbf{ \large
\section{\label{sec:SI_scaling} $\lambda$-scaling}
}
\end{center}

In the main text, the total rate for the chemical reaction Li~+~NaLi($v=0,\,j=0$) $\to$ Li$_2$~+~Na is obtained by adding together the even and odd contributions due to the exchange symmetry of identical Li nuclei. 
This addition leads to an averaging effect. Indeed, in \cref{figSI:scaling}(a) we see a higher contrast in the reaction rate  for each individual exchange symmetry plotted as a function of $\lambda$. 
We note that each exchange symmetry contribution is a sum of state-to-state contributions, and hence already contains a substantial amount of averaging  as stated in the main text.

\begin{figure}[h!]
\begin{center}
\includegraphics[height=0.22\textheight,keepaspectratio]{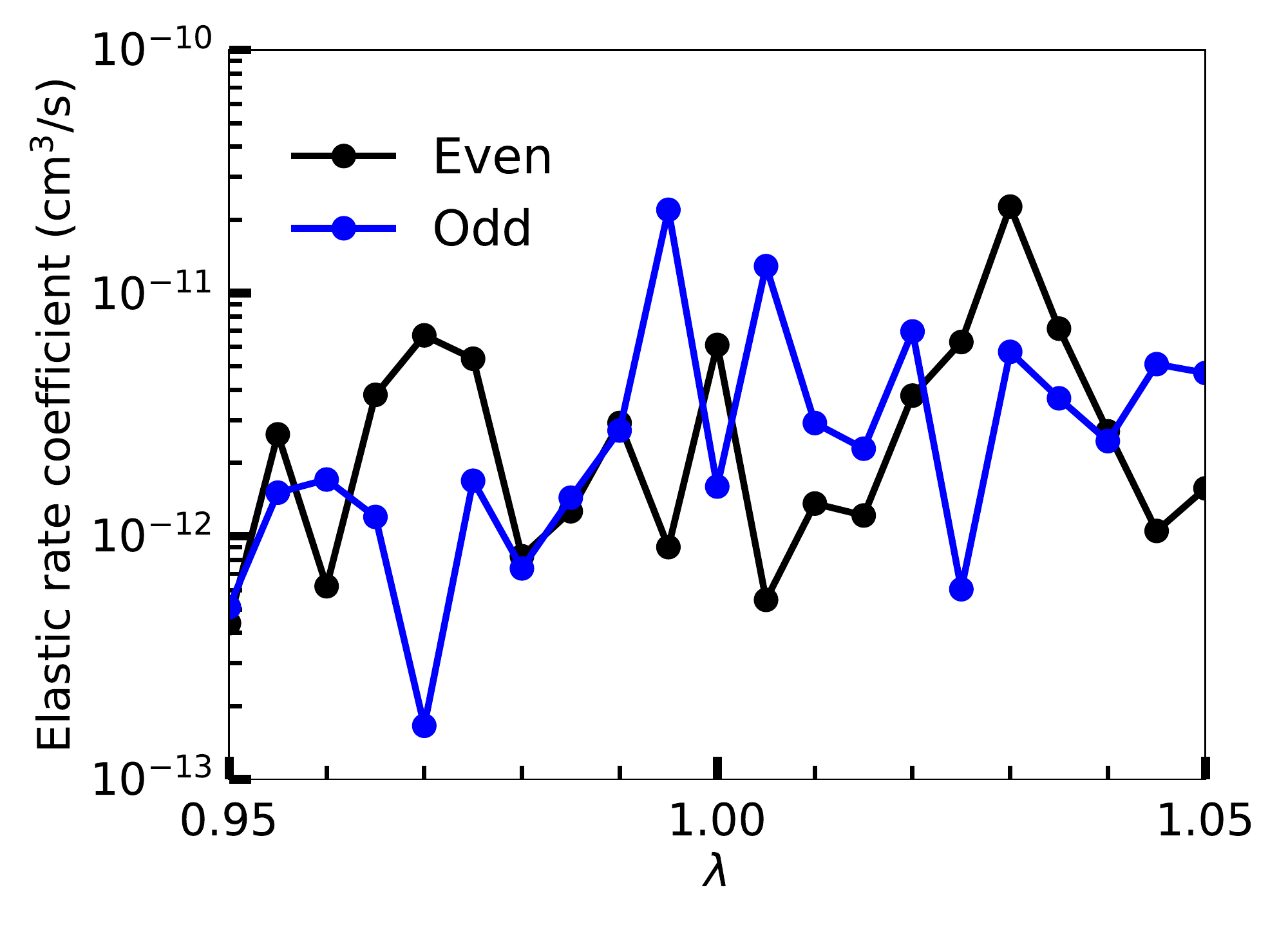}
\includegraphics[height=0.22\textheight,keepaspectratio]{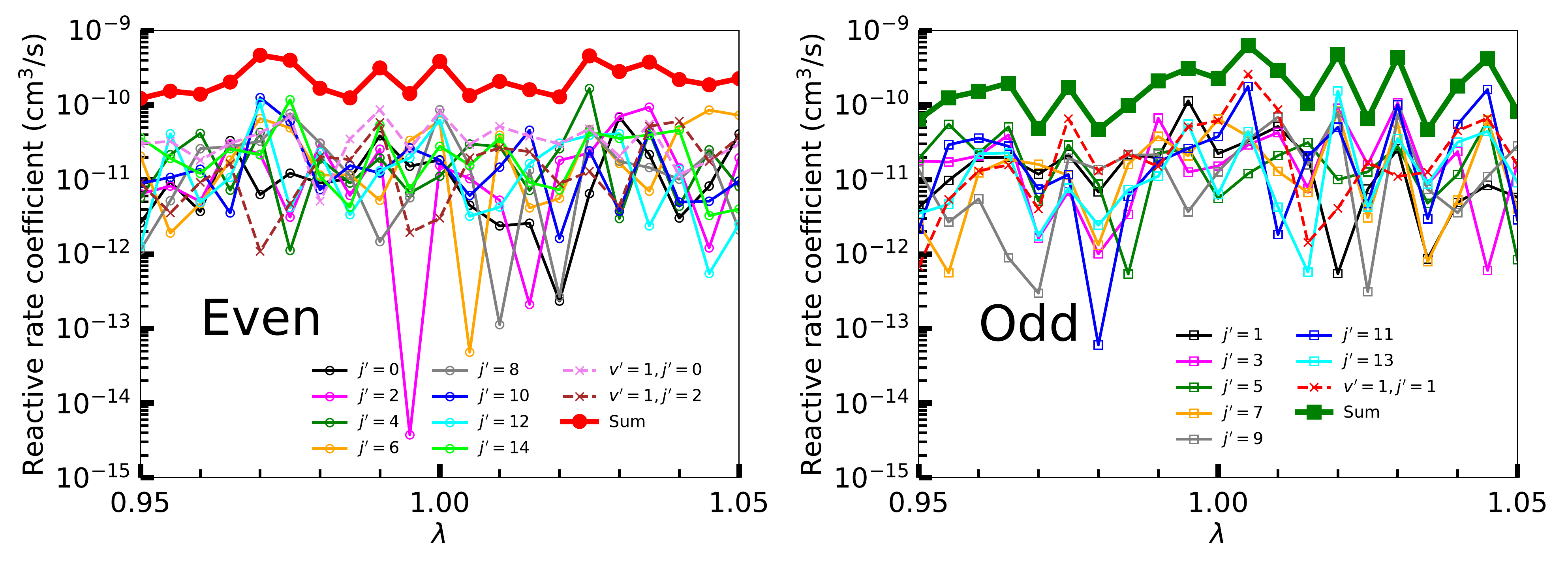}
\end{center}
\caption{
Upper panel: The $\lambda$ dependence of the elastic rate for  Li+NaLi($v=0,\,j=0$) collisions at $E_\mathrm{c}= 1$~$\mu$K.  
Lower panels show state-to-state reaction rates to all  energetically accessible final states ($v^\prime=0,\,j^\prime=0-14$ and $v^\prime=1,\,j^\prime=0-2$) as a function of $\lambda$ at $E_\mathrm{c}= 1$~$\mu$K for even (left) and odd (right) exchange symmetries. The vibrational quantum number $v'=0$ is omitted in the legends. The statistical factors of 1/3 and 2/3 are not applied.
}
\label{figSI:scaling_s-to-s}
\end{figure}

In panel \cref{figSI:scaling}(b), we show the behavior of the state-to-state reaction rates as functions of $\lambda$ for several  product states ($v^\prime,\,j^\prime$) of even exchange symmetry and $J=0^+$. 
We observe a very pronounced $\lambda$-dependence of the state-to-state rates, which can vary by several orders of magnitude  over a narrow interval of $\lambda$ ($\pm1$\%).
%If we increase the $\lambda$ resolution, the range of the maximum and minimum rates will likely increase.
This shows that state-to-state collision dynamics are profoundly affected by the detailed short-range behavior of the PES.

%We show more detailed information about the $\lambda$-scaling calculations because the origin of the sensitivity of the rates to the value of $\lambda$ is an important subject. 
In \cref{figSI:scaling_s-to-s}, we show the elastic rates (upper panel) and all possible state-to-state reaction rates (lower panels) as a functions of $\lambda$.
The elastic rate varies by several orders of magnitude as a function of $\lambda$  for each exchange symmetry. 
While the variations of the elastic  and state-to-state reaction rates might have a common origin  (such as near-threshold resonances), a more detailed study using finer $\lambda$ grids is required to establish the existence of a correlation between the elastic and state-to-state reaction rates.
The total reaction rates for both exchange symmetries vary strongly with $\lambda$  due to the incomplete averaging of the state-to-state rates. 
%The observed significant $\lambda$ dependence of the state-to-state rates can be attribute to near threshold resonances while further detailed calculations are required using finer $\lambda$ grids.

\begin{center}
\textbf{ \large
\section{\label{sec:results_2body} Results with pairwise PES}
}
\end{center}

\begin{figure}[th!]
\begin{center}
\includegraphics[height=0.4\textheight,keepaspectratio]{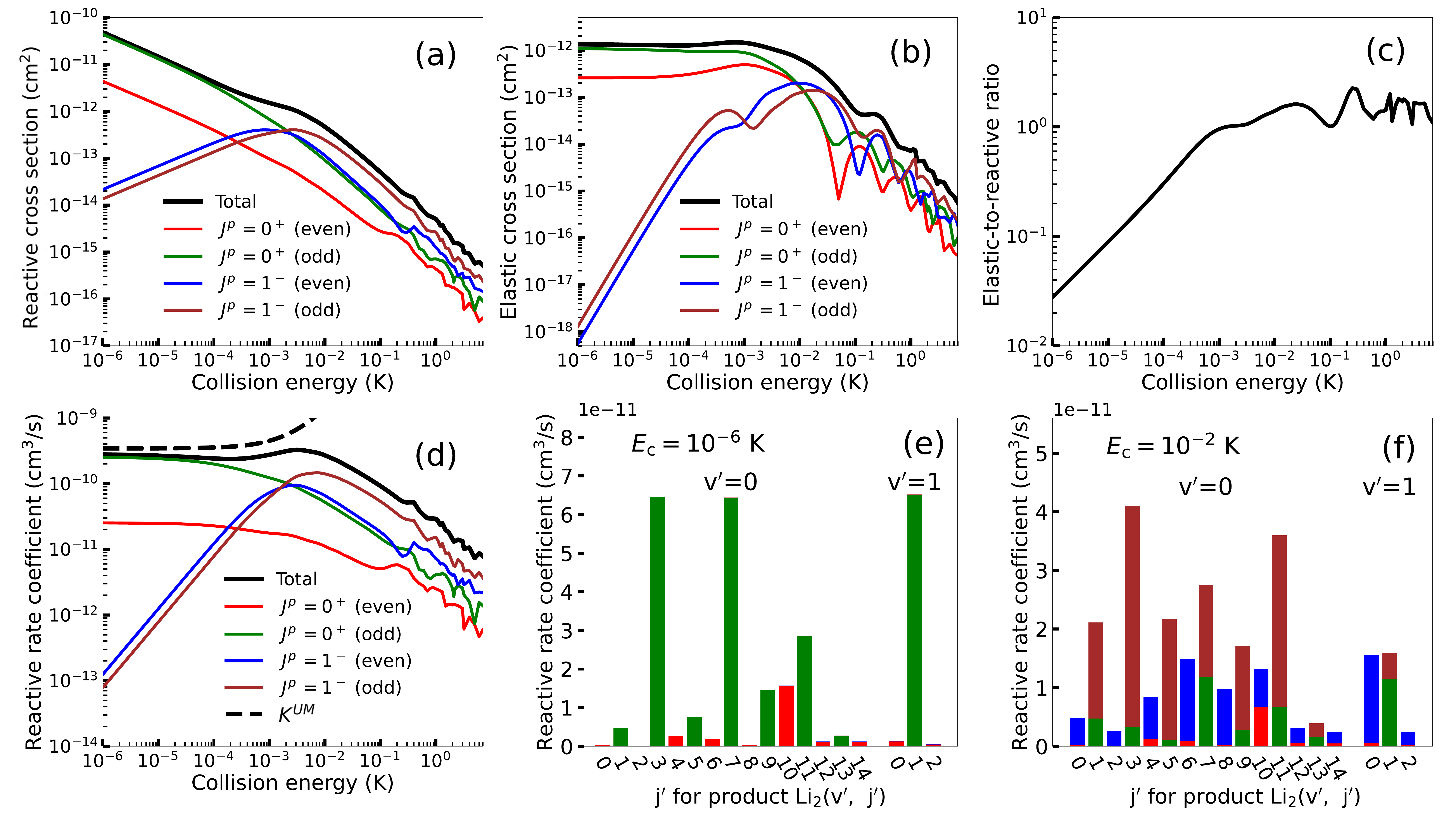}
\end{center}
\caption{
Reactive (a) and elastic (b) cross sections for the  ultracold chemical reaction Li~+~NaLi($v=0,j=0$) $\to$ Li$_2$~+~Na calculated on the pairwise PES (\cref{eq:2body})  and plotted as a function of collision energy for different values of $J$, parity $p$ and exchange symmetry.  
The total integral cross sections (solid black lines) are obtained by summing together the cross sections for even and odd $^6$Li exchange symmetries multiplied by the statistical factors 1/3 and 2/3.
We note that these factors are opposite to those used in previous calculations on the ground-state PESs \cite{Kendrick:2018_Nonad, Makrides:15, Kendrick:21b} because the  $X^1\Sigma^+_g$ and $a^3\Sigma^+_u$ electronic states of $^6$Li$_2$ are even and odd with respect  to the exchange of identical Li nuclei.
(c) The ratio of elastic to reactive cross sections $\gamma$ as a function of collision energy. 
(d) Reaction rate as a function of collision energy. The universal rate is shown by the black dashed line.
Nascent product state distributions of Li$_2$ over final rovibrational states ($v',j'$) at  $E=1$ $\mu$K  (e) and $E=10$~mK (f). The contributions of each $J$,  $p$, and exchange symmetry are color coded in the same way as in panels (a), (b), and (d).
}
\label{figSI:2-body}
\end{figure}

\begin{figure}[bt!]
\begin{center}
\includegraphics[height=0.4\textheight,keepaspectratio]{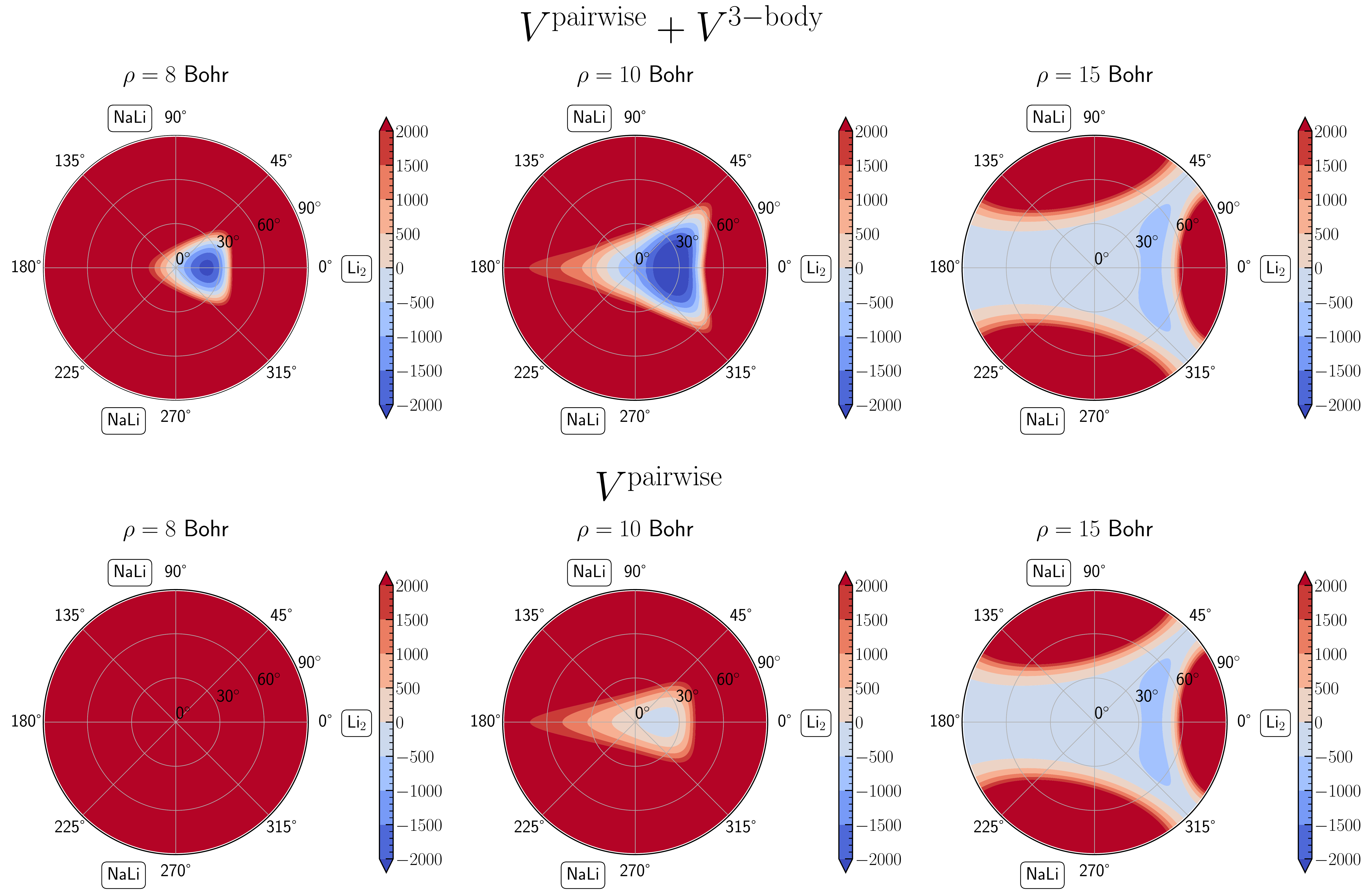}
\end{center}
\caption{
Comparison of the pairwise (bottom panel) and full (top panel) PESs of the NaLi$_2$ reaction complex. 
}
\label{fig:2boby_vs3body}
\end{figure}

To explore the effect of three-body interactions on ultracold reaction observables, we performed reactive scattering computations on the two-body pairwise PES given by \cref{eq:2body}.  Note that the pairwise PES has the same long-range behavior as the full PES given by \cref{eq:Vfull}. Thus, the only difference between these two PESs comes from three-body interactions at short range, as shown in Fig.~\ref{fig:2boby_vs3body}.

 %of the results from those calculated with the full PES (see the main text) is unambiguously due to the difference in the short-range behavior of the PES because 

\Cref{figSI:2-body} shows the reactive and elastic cross sections and their ratio calculated using the pairwise PES (\cref{eq:2body}).
%  with $\lambda=0$.  
We observe that the contribution of the odd exchange symmetry for the identical Li nuclei (green) is almost 10 times larger than that of the even exchange symmetry in the $s$-wave regime for the reactive cross sections. 
On the other hand, we see comparable contributions from both exchange symmetries  in the full PES results with $\lambda=1$ (see the main text). This implies that short-range three-body interactions can have a significant effect on  the dependence of reactive scattering cross sections on the exchange symmetry.
%The dependence comes from a subtle difference of the wave functions with different symmetries, it is difficult to predict the exchange symmetry dependence from the shape of the PES.
The large difference between the exchange symmetries is most likely due to quantum interference  between the Li exchange and non-exchange processes as observed for H~+~HD $\to$ H~+~HD \cite{Croft:17c}, although it is difficult  to determine the extent of the interference based on the shape of the PES.

\Cref{figSI:2-body}(c) shows that the reactive cross section is much larger than the elastic cross section at ultralow energies.
The  total reaction rate ($2.78 \times 10^{-10}$ cm$^3$/s) calculated on the pairwise PES agrees well with the universal rate ($3.45 \times 10^{-10}$ cm$^3$/s) at $E=1$ $\mu$K as shown in \cref{figSI:2-body} (d) and is comparable with the rate ($2.83 \times 10^{-10}$ cm$^3$/s)) calculated with the full PES. 
However, the numerical agreement of the total reaction rates between  the pairwise PES ($\lambda=0$) and full PES ($\lambda=1$) is likely the result of a coincidence caused by rapid oscillations in the total rate as a function of $\lambda$ as shown in Fig.~3(b) in the main text and \cref{figSI:scaling,figSI:scaling_s-to-s}.
The product state distributions calculated for the pairwise and full PESs are very different as shown in \cref{figSI:2-body}(e). Therefore, despite the coincidental agreement between the $\lambda=0$ and $\lambda=1$ total reaction rates,  the three-body term provides a significant contribution to short-range reaction dynamics.

Finally, we observe that  the total even and odd symmetry rates vary strongly with $\lambda$ in the vicinity of $\lambda=1$ differing by as much as a factor of 6. This is similar to the $\lambda=0$ case, where the odd exchange symmetry contributes almost 10 times more than the even exchange symmetry. While the  $\lambda=0$ case does not correspond to a physical interaction PES, the factor of 10 difference between the even and odd exchange symmetry contributions could constitute an interesting dynamical feature of the pairwise interaction PES, which deserves further study. 
%This is now mentioned on p.~S8 of the SI.

% As discussed above, $\lambda=0$ and $\lambda=1$ have totally different contributions of the exchange symmetry components and state-to-state components, thus the short-range dynamics on these two PESs are strongly  different. }
%The density of states of the collision complex on the pairwise PES is smaller than that on the full PES. This indicates that the effect of  exchange symmetry of identical Li nuclei  on  the  $s$-wave reaction rate  is more important than that of the density of states. 

\Cref{figSI:2-body} (e) shows the distributions of Li$_2$ reaction products over their final rovibrational states ($v^\prime, j^\prime$) at $E=1$~$\mu$K calculated on the pairwise PES.
The contribution from even exchange symmetry with $J^p=0^+$ (green bars) is dominant, and we observe highly non-uniform distributions in each exchange symmery, with enhanced  reaction rates for the production of Li$_2$ molecules in the final rovibrational states $j^\prime=3$ and $7$ for $v^\prime=0$ and $j^\prime=1$  for $v^\prime=1$.
As shown in panel (f), $J\ge 1$ contributions exceed those from $J=0$ at $E=10$ mK for  most of the final states. 
We also observe a decrease in product state selectivity compared to the $s$-wave limit because the same final rotational states can now be populated in collisions occurring at multiple $J$ and $l$.\\
%We note that the decrease of selectively comes from the averaging the contributions of independent partial waves for different $J$.
% rather than any specific dynamical feature.
\newline

\bibliography{cold_mol_new}